\pdfoutput=1
\RequirePackage{ifpdf}
\ifpdf % We are running pdfTeX in pdf mode
\documentclass[pdftex]{sigma}
\else
\documentclass{sigma}
\fi

\def\sign#1{{\,\operatorname{sign}(#1)}}

\numberwithin{equation}{section}

\begin{document}

\newcommand{\arXivNumber}{1504.01953}

%\allowdisplaybreaks

\renewcommand{\thefootnote}{$\star$}

\renewcommand{\PaperNumber}{080}

\FirstPageHeading

\ShortArticleName{Structure Preserving Discretizations of the Liouville Equation and their Numerical Tests}

\ArticleName{Structure Preserving Discretizations\\ of the Liouville Equation and their Numerical Tests\footnote{This paper is a~contribution to the Special Issue on Exact Solvability and Symmetry Avatars
in honour of Luc Vinet.
The full collection is available at
\href{http://www.emis.de/journals/SIGMA/ESSA2014.html}{http://www.emis.de/journals/SIGMA/ESSA2014.html}}}

\Author{Decio LEVI~$^\dag$, Luigi MARTINA~$^\ddag$ and Pavel WINTERNITZ~$^{\dag\S}$}

\AuthorNameForHeading{D.~Levi, L.~Martina and P.~Winternitz}

\Address{$^\dag$~Mathematics and Physics Department, Roma Tre University
and Sezione INFN of Roma Tre,\\
\hphantom{$^\dag$}~Via della Vasca Navale 84, I-00146 Roma, Italy}
\EmailD{\href{mailto:decio.levi@roma3.infn.it}{decio.levi@roma3.infn.it}}
%\URLaddressD{\url{http://www.matfis.uniroma3.it/persone/docenti/docenti_beige.php?persona=112}}

\Address{$^\ddag$~Dipartimento di
Matematica e Fisica~- Universit\`a del Salento and Sezione INFN of
Lecce,\\
\hphantom{$^\ddag$}~Via per Arnesano, C.P.~193 I-73100 Lecce, Italy}
\EmailD{\href{mailto:luigi.martina@le.infn.it}{luigi.martina@le.infn.it}}

\Address{$^\S$~D\'epartement de math\'ematiques et de statistique and Centre de recherches math\'ematiques,\\
\hphantom{$^\S$}~Universit\'e de Montr\'eal, C.P.~6128, succ. Centre-ville, Montr\'eal (QC) H3C 3J7,\\
\hphantom{$^\S$}~Canada (permanent address)}
\EmailD{\href{mailto:wintern@crm.umontreal.ca}{wintern@crm.umontreal.ca}}

\ArticleDates{Received April 01, 2015, in f\/inal form September 22, 2015; Published online October 02, 2015}

\Abstract{The main purpose of this article is to show how  symmetry structures   in
partial dif\/ferential equations can be preserved in a discrete world and
ref\/lected in dif\/ference schemes.
    Three dif\/ferent structure preserving discretizations of the Liouville
equation are presented and then used to solve specif\/ic boundary
value problems. The results are compared with exact solutions
satisfying the same boundary conditions. All three discretizations are
on four point lattices.
One preserves linearizability of the equation, another the inf\/inite-dimensional symmetry group as higher symmetries, the third one preserves the
maximal f\/inite-dimensional subgroup of the symmetry group as point
symmetries.
A 9-point invariant scheme that gives a better
approximation of the equation,
but signif\/icantly worse numerical results for solutions is presented and discussed.}


\Classification{17B80; 22E60; 39A14; 65Mxx}

\rightline{\it Dedicated to Luc Vinet on the occasion of his 60th birthday.}

\renewcommand{\thefootnote}{\arabic{footnote}}
\setcounter{footnote}{0}

\vspace{-2mm}

\section{Introduction}

This article is part of a general program the aim of which is to make full use of the theory of Lie groups to study the solution space of discrete equations and in particular to solve dif\/ference equations~\cite{12,13,h14,LOTW,LW91,14,LWY2002,15}. This is one of the areas to which Luc Vinet made important contributions \cite{9,8,10, 7}.

S.~Lie introduced what is now called Lie groups as groups of transformations of the independent and dependent variables f\/iguring in a system of dif\/ferential equations \cite{Lie,Olver}. Of special importance are symmetry groups, transforming solutions into solutions. These may be point transformations, where new variables depend only on the old ones. They may be contact transformations, where the new variables depend also on the f\/irst derivatives of the dependent variables. They may also be generalized symmetries where the new variables can also depend on all derivatives of the old ones.

Lie's method is particularly powerful for ordinary dif\/ferential equations (ODEs). A one-dimensional point symmetry group can be used to lower the order of the ODE by one. An $n$-dimensional (solvable) Lie point symmetry group can be used to decrease the order by $n$. Thus, if the ODE is of order $n' \le n$ Lie group theory can provide the general solution (i.e., one that satisf\/ies arbitrary initial conditions) in explicit or implicit analytic form. For partial dif\/ferential equations (PDEs)  the Lie point symmetry group is used to decrease the number of independent variables in the equation and to provide special solutions ({\it group invariant solutions}), satisfying particularly  symmetrical boundary conditions.

The aim of this general program is to extend the use of Lie symmetry groups to dif\/ference systems ($\Delta$S), i.e., to dif\/ference equations together with the lattice they are written on.

The program has two complementary aspects, an analytical and a numerical one.

The aim of the {\it analytical aspect} is to determine the maximal symmetry group of the $\Delta$S, i.e., the group of transformations that takes solutions into solutions, and then to use it to obtain exact analytic solutions, at least special ones, if possible general ones.  The $\Delta$S to which the approach is applied can come from the study of discrete physical, chemical, biological or other systems, for which symmetries play an important role. Among them we mention phenomena in crystals, or in atomic or molecular chains.

 On the other hand $\Delta$S can be obtained by discretizing ODEs, or PDEs, that have nontrivial symmetry groups ref\/lecting fundamental physical laws such as Galilei, Lorentz, or conformal invariance.  At the scale of the Planck length space-time may very well be discrete. In this case continuous equations are approximations (continuous limits) of discrete ones.  From the physical point of view the symmetries are very important and should be preserved, e.g., when studying quantum f\/ield theories on lattices.

One way of preserving symmetries in a discretization of continuous equations (the one used in this article) is to use symmetry adapted lattices that themselves transform under the group action. This greatly enlarges the set of equations for which symmetry preserving discretization is possible.  We will however see that in some cases only a subgroup of the Lie point symmetry group can be preserved as point symmetries.

The {\it numerical aspect} of our program is the following. When solving an ODE or PDE nu\-merically it is always necessary to replace the continuous equation by a dif\/ference system. This can be done in a standard manner, applicable to all equations, simply by replacing derivatives by discrete derivatives. The other possibility takes us directly into the f\/ield of geometric integration~\cite{geo1,geo2,geo3,geo4}. The idea is to focus on some important feature of the underlying problem and to preserve it in the discretization. Such a feature may be, for instance linearizability, hamiltonian structure, integrability in the sense of the existence of a Lax pairs and generalized symmetries or point and contact symmetries. We are concentrating on point symmetries and exploring the possibility and usefulness of including them in numerical calculations.

Earlier work has shown that for f\/irst-order ODEs preserving a one-dimensional symmetry group provides an exact discretization~\cite{RW}. For second-order ODEs preserving a 3-dimensional symmetry group often provides analytically solvable schemes (either via a Lagrangian~\cite{DKW,DKW1} or via the adjoint equation method~\cite{DKKW}). For third- and higher-order ODEs symmetry preserving discretization provides numerical solutions that are, usually, closer to exact ones then those obtained by other methods, specially near to the singularities~\cite{BCGW,RebW}. For previous work on PDEs see \cite{biblio2,bihlo,biblio4,BuDo,DoKo,grant,LeRo,LeRo1,ReVa,2,VaWi}.

Several recent articles \cite{1,3,2} were devoted to discretizations of  the Liouville equation \cite{4}
\begin{gather}
 z_{x y} = e^z \label{1.1},
\end{gather}
or its algebraic version
\begin{gather}
u u_{x y} -u_x u_y = u^3 , \qquad u = e^z. \label{1.2}
\end{gather}

The Liouville equation is of interest for many reasons. In dif\/ferential geometry it is the equation satisf\/ied by the conformal factor $z(x,y)$ of the metric $ds^2=z^2 (dx^2+dy^2)$ of a two-dimensional space of constant curvature~\cite{5}.
 In the theory of inf\/inite-dimensional nonlinear integrable systems it is the prototype of a nonlinear partial dif\/ferential equation (PDE) linearizable by a transformation of variables, involving the dependent variables (and their f\/irst derivatives) alone~\cite{4}
 \begin{gather} \label{1.3}
 u= 2 \frac{\phi_x \phi_y}{\phi^2}, \qquad \phi_{xy}=0.
 \end{gather}
 In Lie theory this is probably the simplest PDE that has an inf\/inite-dimensional Lie point symmetry group \cite{6}. The symmetry algebra of the algebraic Liouville equation~(\ref{1.2}) is given by the vector f\/ields
 \begin{gather} \label{1.4}
 X(f(x)) = f(x) \partial_x - f_x(x) u \partial_u, \qquad Y(g(y))=g(y) \partial_y - g_y(y) u \partial_u,
 \end{gather}
 where $f(x)$ and $g(y)$ are arbitrary smooth functions.

 Equation~(\ref{1.4}) is a standard realization of the direct product of two centerless Virasoro algebras and we shall denote the corresponding Lie group $ {\rm VIR}(x) \otimes {\rm VIR}(y)$.  Restricting~$f(x)$ and~$g(y)$ to second-order polynomials we obtain the maximal f\/inite-dimensional subalgebra ${\mathfrak{sl}}_x (2 , \mathbb{R})\bigoplus {\mathfrak{sl}}_y (2 , \mathbb{R})$ and the corresponding f\/inite-dimensional subgroup $ {\rm SL}_x (2 , \mathbb{R}) \otimes {\rm SL}_y (2 , \mathbb{R})$ of the symmetry group.

 The Liouville equation is also an excellent tool for testing numerical methods for solving PDE's, since equation~(\ref{1.3}) provides a very large class of exact analytic solutions, obtained by putting
 \begin{gather} \label{1.5}
 \phi(x,y)=\phi_1(x)+\phi_2(y),
 \end{gather}
 where $\phi_1(x)$ and $\phi_2(x)$ are arbitrary $\mathbb C^{(2)}(I)$ functions on some interval~$I$.

 In \cite{1} Adler and Startsev presented a discrete Liouville equation that preserves the pro\-per\-ty of being linearizable and exactly solvable. In~\cite{2} Rebelo and Valiquette wrote a~discrete Liou\-ville equation that has the same inf\/inite-dimensional ${\rm VIR}(x)\otimes {\rm VIR}(y)$ symmetry group as the continuous Liouville equation.  The transformations are however generalized symmetries, rather than point ones.  In our article~\cite{3} we presented a discretization on a four-point stencil that preserves the maximal f\/inite-dimensional subgroup of the ${\rm VIR}(x) \otimes {\rm VIR}(y)$ group as point symmetries. It was also shown that it is not possible to conserve the entire  inf\/inite-dimensional Lie group of the Liouville equation as {\it point} symmetries. In~\cite{3} we also compared numerical solutions obtained using standard (non invariant) discretizations, the Rebelo--Valiquette invariant discretization~\cite{2} and our discretization with exact solutions (for 3 dif\/ferent specif\/ic solutions). It turned out that the discretization based on preserving the maximal subgroup of point transformations always gave the most accurate results for the considered solutions (all of them strictly positive in the area of integration).

The purpose of this article is to further explore and compare the dif\/ferent discretizations of the Liouville equation from two points of view. One is a theoretical one, namely to investigate the degree to which dif\/ferent discretizations preserve the qualitative feature of the equation: its exact linearizability, its inf\/inite-dimensional Lie point symmetry algebra, the behavior of the zeroes of the solutions.  The other point of view is that of geometric integration: what are the advantages and disadvantages of the dif\/ferent discretizations as tools for obtaining numerical solutions.

In Section~\ref{section2} we reproduce our previous \cite{3}  ${\rm SL}_x (2, \mathbb{R}) \otimes {\rm SL}_y (2, \mathbb{R})$ symmetry preserving discretization using a 4-point stencil and show that after a slight modif\/ication it can reproduce solutions that have horizontal or vertical lines of zeroes (or both). In Section~\ref{section3} we propose an alternative discretization, using a 9-point stencil, instead of the 4-point one. It approximates  the continuous Liouville equation with $\epsilon^2$ precision, as opposed to the $\epsilon$ precision of the 4-point discretization.  We show that increasing the number of points does not allow us to preserve the entire inf\/inite-dimensional symmetry algebra, nor to treat the lines of zeroes of solutions in a~satisfactory manner.
 Further, in Section~\ref{section3} we take a specif\/ic exact solution of the
continuous algebraic Liouville equation~\eqref{1.2} and approximate it on a~9-point lattice by a numerical solution.
 The Adler--Startsev discretization~\cite{1} is reproduced in Section~\ref{section4} in a form suitable for numerical calculations. Section~\ref{section5} is devoted to numerical tests of the invariant 4-point
scheme. Five dif\/ferent exact solution of the algebraic Liouville
equation are presented and then used to calculate boundary conditions on
two lines parallel to the~$x$ and~$y$ coordinate axes, respectively. The solutions are then calculated numerically using four dif\/ferent discretizations. We compare the validity of the dif\/ferent methods and their qualitative features.
Some general
conclusions, placing this article in the context of geometric
integration, are presented in the f\/inal Section~\ref{section6}.

\section[Point symmetries on a four point lattice and solutions with zeroes]{Point symmetries on a four point lattice\\ and solutions with zeroes}\label{section2}

In our previous article \cite{3} we discretized the algebraic Liouville equation (\ref{1.2}) on a four point regular orthogonal lattice preserving the $ {\rm SL}_x (2 , \mathbb{R})\otimes {\rm SL}_y (2 , \mathbb{R})$ subgroup of its Lie point symmetries.  The discretization was shown to provide good numerical results for solutions that were strictly positive in the entire integration  region (a quadrant to the right and above a chosen point ($x_0,y_0$), i.e., for $x \ge x_0$, $y \ge y_0$).

A particular property of the Liouville equation is that the zeroes of its solutions are not isolated. They occur on lines parallel to the~$x$ or~$y$ axes. Indeed, consider the inf\/inite family of solutions of~(\ref{1.2}) parametrized by two arbitrary smooth functions of one variable~$\phi_1(x)$, $\phi_2(y)$~(\ref{1.5}). We take a region in which we have $\phi_1(x)+\phi_2(y) \ne 0$. Zeroes of $u(x,y)$ occur if~$\phi_{1,x}(x)$, or $\phi_{2,y}(y)$  are zero at some point~$x_s$, or~$y_s$ (or both), respectively. We then have
\begin{gather} \label{a2.2}
u(x_s,y)=0, \quad \forall\, y, \qquad \mbox{or} \qquad u(x,y_s)=0, \quad \forall\, x .
\end{gather}
This must be ref\/lected in any computational scheme and the value $u(x,y)=0$ will also occur on the intersection with the corresponding coordinate axis.

In \cite{3} we considered several dif\/ferent boundary value problems. Here we restrict to the case of boundary conditions given on the lines $x \ge x_0$, $y \ge y_0$ parallel to the coordinate axes. We can impose
\begin{gather*} %\label{a2.3}
u(x_s,0)=0  \qquad \mbox{and/or} \qquad u(0,y_s)=0
\end{gather*}
in order to obtain a solution satisfying (\ref{a2.2}).

The ${\rm SL}_x (2 , \mathbb{R})\otimes {\rm SL}_y (2 , \mathbb{R})$ invariants used in~\cite{3} to describe both the lattice and the discrete algebraic Liouville equation  on a four point stencil were
\begin{gather}
 \xi_1=  \frac{\left(x_{m,n+1}-x_{m,n}\right)
   \left(x_{m+1,n+1}-x_{m+1,n}\right)}{\left(x_{m,n}-x_{m+1,n}\right)
   \left(x_{m,n+1}-x_{m+1,n+1}\right)} ,  \nonumber\\
 \eta_1=  \frac{\left(y_{m,n}-y_{m+1,n}\right)
   \left(y_{m,n+1}-y_{m+1,n+1}\right)}{\left(y_{m,n+1}-y_{m,n}\right)
   \left(y_{m+1,n+1}-y_{m+1,n}\right)}, \label{a2.4} \\
 J_1= u_{m+1,n} u_{m,n+1} h^2 k^2, \qquad
 J_2 =  u_{m,n} u_{m+1,n+1} h^2 k^2. \label{a2.5}
\end{gather}
The lattice equations
\begin{gather} \label{a2.6}
\xi_1=0, \qquad \eta_1=0
\end{gather}
are satisf\/ied by the uniform orthogonal lattice
\begin{gather} \label{a2.7}
x_{m,n}=hm+x_0, \qquad y_{m,n}=kn+y_0,
\end{gather}
where the scale factors $h$ and $k$ are the same as in (\ref{a2.5}). The continuous limit corresponds to $h \rightarrow 0$, $k \rightarrow 0$.
Two further independent  $ {\rm SL}_x (2 , \mathbb{R}) \otimes {\rm SL}_y (2 , \mathbb{R})$ invariants exist on the four point stencil but any combination of them will either vanish, or be inf\/inite on the lattice given by~(\ref{a2.6}) (see~\cite{3}).

The Liouville equation~(\ref{1.2}) was approximated in \cite{3} by the dif\/ference scheme
\begin{gather}
 J_2-J_1=a  \sign{J_1}|J_1|^{3/2}+b   \sign{J_2}J_1 |J_2|^{1/2}\nonumber\\
 \hphantom{J_2-J_1=}{} +c \sign{J_1} |J_1|^{1/2}J_2+d \sign{J_2}|J_2|^{3/2}, \label{a2.8}\\
\nonumber
  \xi_1=\eta_1=0, \qquad a+b+c+d=1.
\end{gather}
The symbols $\sign{J_1}$ and $\sign{J_2}$ were omitted in~\cite{3} and were not necessary as we restricted our formulation to strictly positive solutions.
Equation~(\ref{a2.8}) can be solved for $u_{m+1,n+1}$ in terms of $u_{n,m}$,  $u_{n+1,m}$ and  $u_{n,m+1}$.
On the f\/irst stencil we have $m=n=0$. The boundary conditions are $u_{m,0}=f(m)$ and $u_{0,n}=g(n)$ with~$f$ and~$g$ given.

Let us now rewrite the recurrence relation~(\ref{a2.8}) in terms of $u_{m,n}$, choose $b=d=0$, $c=1-a$, $a \in \mathbb R$ (in order to have an explicit  scheme) and solve for $u_{m+1,n+1}$. We have
\begin{gather} \label{a2.9}
u_{m+1,n+1} =  \frac{u_{m,n+1} u_{m+1,n}}{u_{m,n}} A_{m,n+1;m+1,n}, \\
\label{a2.10}
A_{m,n+1;m+1,n}=\frac{1+ah k  \sign{u_{m+1,n} u_{m,n+1} }\sqrt{|u_{m,n+1} u_{m+1,n}|}}{1+(a-1) h k  \sign{u_{m+1,n} u_{m,n+1}}\sqrt{|u_{m,n+1} u_{m+1,n}|}}.
\end{gather}
The expression $\sign{u_{m+1,n} u_{m,n+1}}$ follows from~(\ref{a2.8}). Here the {\it sign} before the square root is important since it will change when the sign of $u$ changes in the recurrence relation.

We shall use  (\ref{a2.9}), (\ref{a2.10}) to investigate the behaviour of the numerical schemes for solutions that have rows (horizontal lines) or columns (vertical lines) of zeroes. We impose boundary conditions on the lines $x=x_s$ and $y=y_s$. To see the inf\/luence of the boundary conditions we introduce small quantities $\mu$ and $\nu$ on the coordinate axes that will later be set to zero. We shall see that these small values do not propagate elsewhere but are conf\/ined to the columns and rows where they were introduced. This procedure is analogous to ``singularity conf\/inement''~\cite{gr, grp} used as an integrability criterion for dif\/ference equations.

We f\/irst note that we have
\begin{gather} \label{a2.13}
\lim_{u_{m,n+1} \rightarrow 0}  A_{m,n+1;m+1,n}= \lim_{u_{m+1,n} \rightarrow 0}  A_{m,n+1;m+1,n}=1.
\end{gather}
Three cases will be considered separately:

1.~A column of zeroes.
The boundary conditions are
\begin{gather*} %\label{b2.11}
u_{m_0,0}=\mu, \qquad u_{m,0} \ne 0 \qquad \mbox{for} \quad m \ne m_0.
\end{gather*}
Using (\ref{a2.9}), (\ref{a2.13}) we obtain  expressions for $u_{m_0,n}$, namely
\begin{gather*} %\label{b2.12}
u_{m_0,n}=\frac{u_{m_0-1,n}}{u_{m_0-1,0}} \mu, \qquad n \ge 0.
\end{gather*}
In the column to the right of the zeroes we obtain two equivalent expressions:
\begin{gather} \label{b2.13a}
u_{m_0+1,n} = u_{m_0+1,n-1}\frac{u_{m_0-1,n}}{u_{m_0-1,n-1}}, \\ \label{b2.13b}
u_{m_0+1,n} = u_{m_0-1,n}\frac{u_{m_0+1,0}}{u_{m_0-1,0}}.
\end{gather}
Thus the zero quantity $\mu$ cancels out and $u_{m_0+1,n}$ is f\/inite and nonzero for all $n \ge 0$.  Moreover $u_{m_0+1,n}$ is expressed in terms of the given initial values and values calculated at previous nonzero values.

2.~A row of zeroes can be treated completely analogously.
The boundary conditions are replaced by
\begin{gather*} %\label{b2.14}
u_{0,n_0}=\nu,   \qquad u_{0,n}\ne 0 \qquad \mbox{for}\quad n\ne n_0,
\end{gather*}
and we obtain
\begin{gather*} %\label{b2.15}
u_{m,n_0}=\frac{u_{m,n_0-1}}{u_{0,n_0-1}} \nu, \qquad m \ge 0,
\end{gather*}
i.e., a row of zeroes for $\nu=0$. The row above the zeroes satisf\/ies
\begin{gather} \label{b2.16a}
u_{m,n_0+1}  =  u_{m-1,n_0+1} \frac{u_{m,n_0-1}}{u_{m-1,n_0-1}}, \\ \label{b2.16b}
u_{m,n_0+1}  =  u_{m,n_0-1} \frac{u_{0,n_0+1}}{u_{0,n_0-1}}.
\end{gather}

3.~Two intersecting lines of zeroes.
The boundary conditions are
\begin{gather*} %\label{b2.17}
u_{0,n_0}=\nu, \qquad u_{m_0,0}=\mu; \qquad u_{m,0} \ne 0 \qquad \mbox{for}\quad m\ne m_0, \qquad u_{0,n}\ne 0 \qquad \mbox{for}\quad n\ne n_0.
\end{gather*}
Using the same considerations as above we f\/ind a column and a row of zeroes satisfying
\begin{gather*} %\label{b2.18}
u_{m_0,n} =\frac{u_{m_0-1,n}}{u_{m_0-1,0}} \mu, \qquad n \ne n_0, \qquad
u_{m,n_0}=\frac{u_{m,n_0-1}}{u_{0,n_0-1}} \nu, \qquad m \ne m_0,
\\ %\label{b2.19}
  u_{m_0,n_0}=\frac{u_{m_0-1,n_0-1}}{u_{0,n_0-1} u_{m_0-1,0}} \mu \nu.
\end{gather*}
Thus, for $\mu=0$, $\nu=0$ the solutions $u_{m,n}$ have zeroes precisely where they should.  Now let us use~(\ref{a2.9}),~(\ref{a2.13})  to calculate the values of $u_{m_0+1,n}$ and $u_{m,n_0+1}$, i.e., the column at the right and the row above the zeroes. The f\/inal result is that~(\ref{b2.13a}) is valid for all $n \ne n_0,n_0+1$ and (\ref{b2.16a}) for all  $m \ne m_0, m_0+1$ with
\begin{gather*} %\label{b2.20}
u_{m_0+1,n_0+1}=u_{m_0-1,n_0+1}\frac{u_{m_0+1,n_0-1}}{u_{m_0-1,n_0-1}},
\end{gather*}
while (\ref{b2.13b})  is valid for all $n \ne n_0$ and~(\ref{b2.16b}) for all  $m \ne m_0$.

Finally we see that the zeroes are conf\/ined to the rows and columns determined by a zero in the boundary condition and that the values of~$u_{m,n}$ everywhere else are f\/inite, non zero and determined by the equations~(\ref{a2.9}),~(\ref{a2.10}) and the boundary conditions. In other words the rows and columns of zeroes do not interfere with the integration algorithm. This will be conf\/irmed by numerical calculations in Section~\ref{section5}.

\section[Invariant discretization  of the algebraic Liouville equation using a larger number of points]{Invariant discretization  of the algebraic Liouville equation\\ using a larger number of points}\label{section3}

There are several reasons to increase the number of points on the stencil that we use.

1.~To determine whether the entire $ {\rm VIR}(x)\otimes {\rm VIR}(y)$ symmetry group can be preserved on a larger lattice.

 2.~To determine whether the only other $ {\rm SL}_x (2 , \mathbb{R}) \otimes {\rm SL}_y (2 , \mathbb{R})$ dif\/ferential invariant~\cite{3}, namely
\begin{gather} \label{xyz}
I_2 = \frac{1}{u^6} \big(2 u u_{xx} - 3 u_x^2\big) \big(2 u u_{yy} - 3 u_y^2\big)
\end{gather}
can be invariantly discretized on a larger lattice. Four points are
clearly not suf\/f\/icient to approximate two f\/irst- and three second-order
derivatives.

3.~To approximate the algebraic Liouville equation with a higher degree of accuracy in~$h$ and~$k$ and thus possibly improve the numerical calculations.

In Section~\ref{section2} and in~\cite{3} we have shown that the Liouville equation can be approximated on 4 points. To approximate an arbitrary second-order PDE for a function $u(x,y)$ we need at least 6 points. An invariant discretization may need more than six.

\begin{figure}[t]\centering
\includegraphics[width=0.45\textwidth]{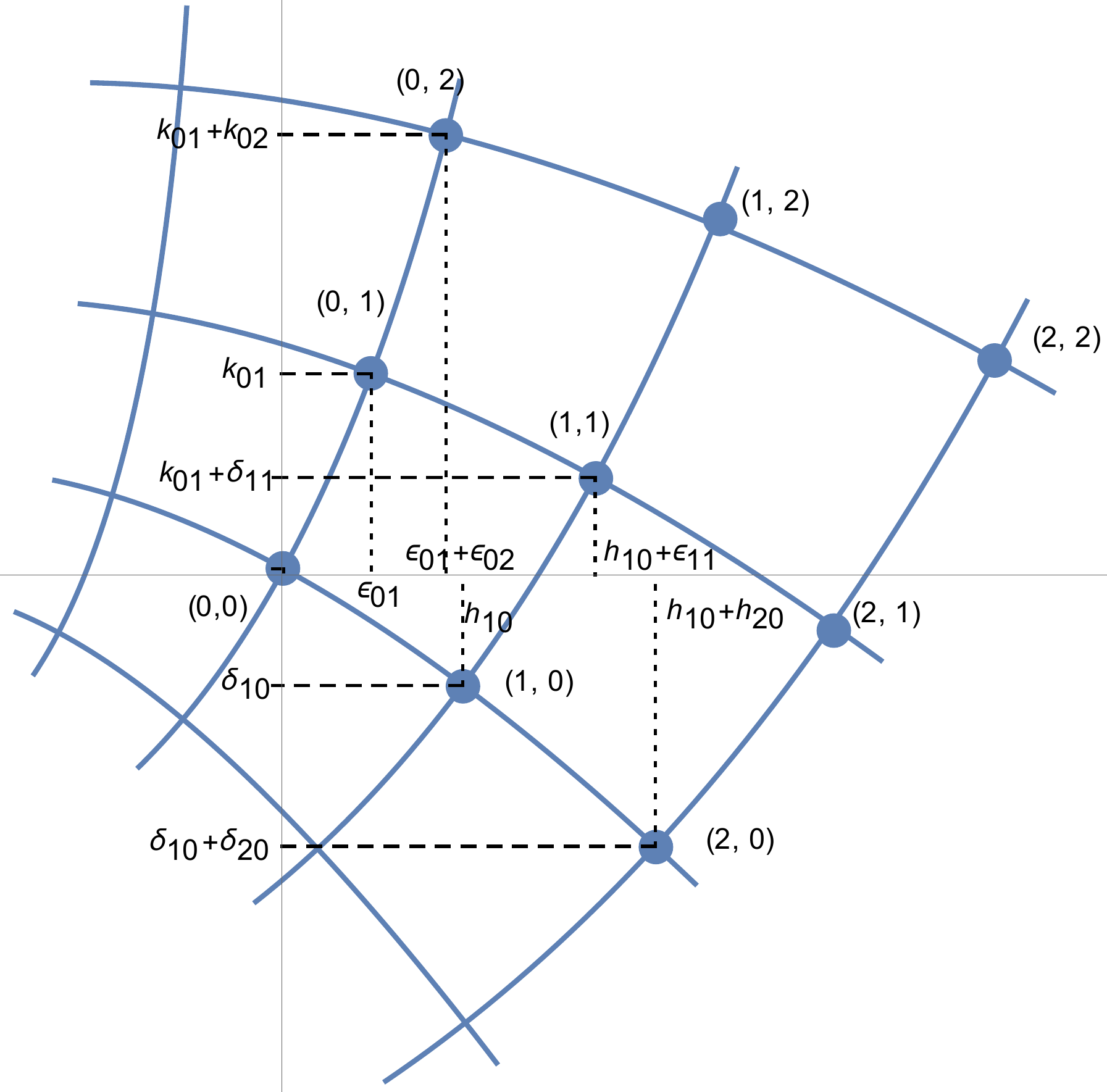}
\caption{Points on a general lattice, e.g.,  $x_{00}=x$, $x_{10}=x+h_{10}$, $x_{01}=x+\epsilon_{01}$, $x_{11}=x+h_{10}+\epsilon_{11}$, $x_{20}=x+h_{10}+h_{20}$, $x_{02}=x+\epsilon_{01}+\epsilon_{02}$, $x_{12}=x+h_{10}+\epsilon_{11}+\epsilon_{12}$, $x_{21}=x+h_{10}+h_{20}+\epsilon_{21}$, $x_{22}=x+h_{10}+h_{20}+\epsilon_{21}+\epsilon_{22}$, $y_{00}=y$, $y_{01}=y+k_{01}$, $y_{10}=y+\delta_{10}$, $y_{11}=y+k_{01}+\delta_{11}$, $y_{02}=y+k_{01}+k_{02}$, $y_{20}=y+\delta_{10}+\delta_{20}$, $y_{12}=y+k_{01}+k_{02}+\delta_{12}$, $y_{21}=y + k_{01}+\delta_{10}+\delta_{20}$, $y_{22}=y+k_{01}+k_{02}+\delta_{12}+\delta_{22}$. }\label{fig1}
\end{figure}
\begin{figure}[t]\centering
\includegraphics[width=0.28\textwidth]{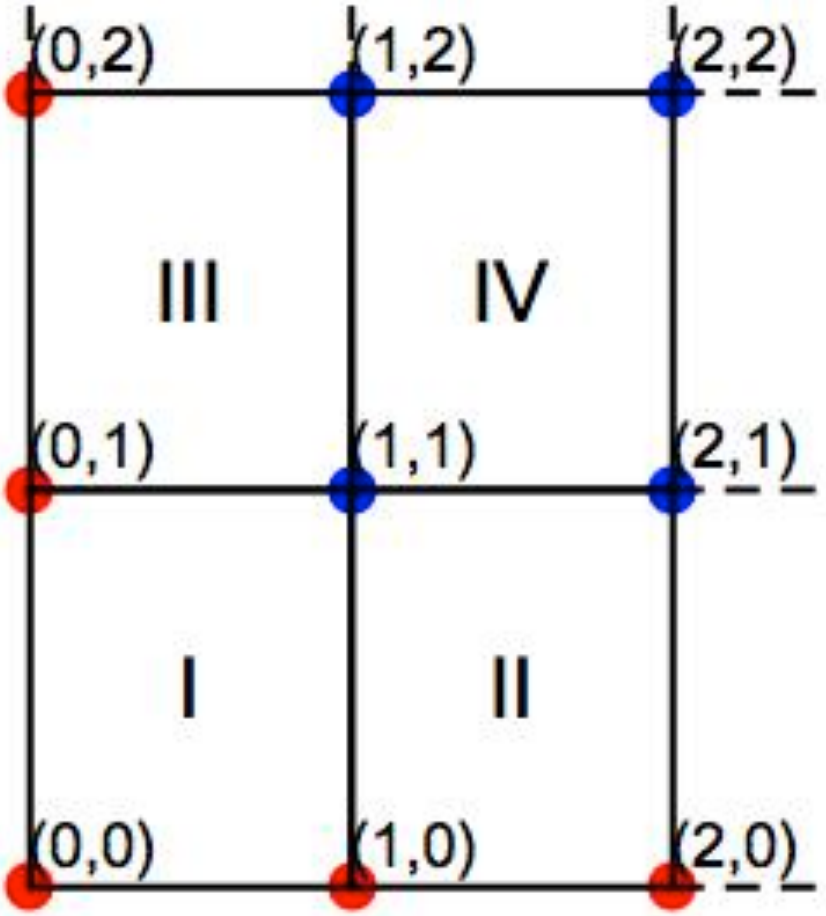}
\caption{A stencil for the 9-points scheme.}\label{f2}
\end{figure}

Equation~(\ref{a2.8}) satisf\/ies
\begin{gather*} %\label{a3.1}
 \lim_{h,k \rightarrow 0} \frac{1}{h^3 k^3} \big\{ J_2-J_1-\big[a |J_1|^{3/2}+b J_1 |J_2|^{1/2}+c|J_1|^{1/2}J_2+d|J_2|^{3/2}\big]\big\}   \\
 \qquad {} =\big[u u_{xy}-u_x u_y -u^3\big][1+\mathcal O(h,k)],
\end{gather*}
and thus provides a f\/irst-order approximation of the algebraic Liouville equation.  In this section we will explore a second-order approximation (order $\mathcal O(h^2,k^2,hk)$) of the equation~(\ref{1.2}). To do this  we
shall use a 9-point stencil as shown on Figs.~\ref{fig1} and~\ref{f2}.
The 4 well-behaved inva\-riants~(\ref{a2.4}),~(\ref{a2.5})   make use of the four vertices of rectangle~I on Fig.~\ref{f2}.
Instead of the vertices of rectangle~I we could use any other 4 points, and we shall use the vertices of the rectangles~II,~III and~IV. The invariants involving the independent variables $\xi_a$ and $\eta_a$ ($a=1, \dots, 4$) all vanish on the orthogonal lattice~(\ref{a2.7}). The invariants depending on the dependent variables~$u_{ij}$ that are f\/inite and nonzero on this lattice  are
\begin{alignat*}{3} %\label{2.6}
&J_1 = u_{01} u_{10} h^2 k^2, \qquad &&J_2 =  u_{00} u_{11} h^2 k^2,&\\
&J_3 = u_{11} u_{20} h^2 k^2, \qquad &&J_4 =  u_{10} u_{21} h^2 k^2,&\\
&J_5 = u_{11} u_{02} h^2 k^2, \qquad &&J_6 =  u_{01} u_{12} h^2 k^2,&\\
&J_7 = u_{12} u_{21} h^2 k^2, \qquad &&J_8 =  u_{11} u_{22} h^2 k^2.&
\end{alignat*}
The quantities $J_1,\dots, J_8$ are {\it linearly} independent but one polynomial relation exists between them, namely
\begin{gather*} %\label{2.7}
J_4 J_6 = J_1 J_7.
\end{gather*}
The continuous limit is obtained by expanding the invariants into Taylor series and then taking $h\rightarrow 0$, $k \rightarrow 0$. We shall assume that they tend to zero at the same rate, i.e., $k=\alpha h$, $\alpha \sim 1$. We have
\begin{gather}
J_1 = h^2 k^2\big[u^2+h u u_x+k u u_y+ h k u_x  u_y +\tfrac{1}{2} h^2 u u_{xx}+\tfrac{1}{2} k^2 u u_{yy}+\cdots\big],\nonumber \\
\nonumber
J_2 = h^2 k^2\big[u^2+h u u_x+k u u_y+ h k u u_{xy}+\tfrac{1}{2} h^2 u u_{xx}+\tfrac{1}{2} k^2 u u_{yy}+\cdots\big], \\ \nonumber
J_3 = h^2 k^2\big[u^2+3h u u_x+k u u_y+\tfrac{1}{2} k^2 u u_{yy}+hk (u u_{xy}+2 u_y u_x)  + h^2 \big(\tfrac{5}{2} u u_{xx}+2 u_x^2\big)+\cdots\big], \\ \nonumber
 J_4 = h^2 k^2\big[u^2+3 h u u_x+k u u_y+\tfrac{1}{2} k^2 u u_{yy}+ hk (2 u u_{xy}+u_y u_x)
 +h^2 \big(\tfrac{5}{2} u u_{xx}+2 u_x^2\big) + \cdots\big], \\ \nonumber
 J_5 = h^2k^2\big[u^2+h u u_x+3 k u u_y+k^2 \big(\tfrac{5}{2} u u_{yy}+2 u_y^2\big)+h k (u u_{xy}+2 u_y u_x)
 +    \tfrac{1}{2} h^2 u u_{xx}+\cdots \big], \\ \nonumber
  J_6 = h^2 k^2 \big[u^2+ h u u_x+3 k u u_y +k^2 \big(\tfrac{5}{2} u u_{yy}+2 u_y^2\big)+h k (2 u u_{xy}+u_y u_x)
   +\tfrac{1}{2} h^2 u u_{xx} + \cdots \big], \\ \nonumber
  J_7 = h^2 k^2 \big[u^2+3 h u u_x+3 k u u_y+k^2 \big(2 u_y^2+\tfrac{5}{2} u u_{yy}\big)+h k (4 u u_{xy}+5 u_y u_x) \\
  \nonumber
\hphantom{J_7 =}{}
   + h^2 \big(2 u_x^2+\tfrac{5}{2} u u_{xx}\big) + \cdots \big], \\ \nonumber
  J_8 = h^2 k^2 \big[u^2+3 h u u_x+3 k u u_y+k^2 \big(\tfrac{5}{2} u u_{yy}+2u_y^2\big)+h k (5 u u_{xy}+4 u_y u_x)  \\
\hphantom{J_8 =}{}   +h^2 \big(\tfrac{5}{2} u u_{xx}+2 u_x^2\big) + \cdots ].\label{2.8}
\end{gather}
We see that $u_{22}$, $u_{02}$, $u_{20}$ and  $u_{00}$ f\/igure only once each in the invariants, namely in $J_8$, $J_5$, $J_3$ and $J_2$, respectively.  On the other hand $u_{01}$, $u_{10}$, $u_{12}$, and $u_{21}$ f\/igure twice each, respectively in ($J_1, J_6$), ($J_1, J_4$), ($J_6, J_7$) and ($J_4,J_7$). The value $u_{11}$ f\/igures in all four of $J_2$, $J_3$, $J_5$ and~$J_8$.

To lowest-order we have
\begin{gather*} %\label{2.9}
J_2-J_1=J_4-J_3=J_6-J_5=J_8-J_7= h^3 k^3 (u u_{xy} - u_x u_y ) [ 1 + \mathcal O(h,k)].
\end{gather*}

To obtain the left hand side of the algebraic Liouville equation (\ref{1.2}) up to order $\mathcal O(h^2,  hk, k^2)$ we need the dif\/ferences $J_{2a}-J_{2a-1}$ to a higher-order than in~(\ref{2.8}), namely
\begin{gather}\nonumber
J_2-J_1  = h^3k^3 \left\{ (u u_{xy}-u_xu_y)+\frac h 2 (u u_{xxy}-u_y u_{xx}) + \frac k 2 (u u_{xyy}-u_x u_{yy}) \right\}, \\ \nonumber
J_4-J_3  = h^3k^3 \left\{ (u u_{xy}-u_xu_y)+\frac {3h} 2 (u u_{xxy}-u_y u_{xx}) + \frac k 2 (u u_{xyy}-u_x u_{yy}) \right\}, \\ \nonumber
J_6-J_5  = h^3k^3 \left\{ (u u_{xy}-u_xu_y)+\frac h 2 (u u_{xxy}-u_y u_{xx}) + \frac {3k} 2 (u u_{xyy}-u_x u_{yy}) \right\}, \\
J_8-J_7  = h^3k^3 \left\{ (u u_{xy}-u_xu_y)+\frac {3h} 2 (u u_{xxy}-u_y u_{xx}) + \frac {3k} 2 (u u_{xyy}-u_x u_{yy}) \right\}.\label{2.8a}
\end{gather}
In \cite{3} equation~(\ref{1.2}) was approximated  to order $\mathcal O(h,k)$. To approximate it to $\mathcal O(h^2, hk, k^2)$ we must get rid of the terms of order $\mathcal O(h,k)$ in~(\ref{2.8a}).

The left hand side is approximated to the needed order by
\begin{gather}
 \alpha [ 4(J_2-J_1)-(J_6-J_5+J_4-J_3)] +\beta [ 4(J_8-J_7)-3(J_6-J_5+J_4-J_3)] \nonumber \\
  \qquad {} = 2 h^3k^3(u u_{xy}-u_xu_y)(\alpha-\beta)\big[1+\mathcal O\big(h^2,   hk,   k^2\big)\big],\label{2.9a}
\end{gather}
where $\alpha$ and $\beta$ are arbitrary real constants.

To express the right-hand side of (\ref{1.2}) we use the basis
\begin{gather}
B_1 = \frac 1 2 (3J_1-J_8)=h^2k^2u^2(1+R_1),\qquad B_2=J_4+J_6-J_8-J_1=h^2k^2R_2, \nonumber\\
B_3 = J_2-J_1=h^2k^2R_3, \qquad B_4=J_4-J_3=h^2k^2R_4,\qquad B_5=J_6-J_5=h^2k^2R_5, \nonumber\\
 B_6 = J_8-J_7=h^2k^2R_6,\label{2.10a}
\end{gather}
where $R_1, \dots, R_6$ are all of the order $\mathcal O(h^2,  hk,  k^2)$.

The left hand side of (\ref{1.2}) is already expressed in this basis (see~(\ref{2.9a}) using $B_3, \dots, B_6$).

From the basis elements (\ref{2.10a}) we can calculate  $u^2$ as
\begin{gather*} %\label{2.11a}
B_1+\sum_{i=2}^6 c_i B_i = h^2k^2u^2\big[1+ \mathcal O\big(h^2,  hk,  k^2\big)\big],
\end{gather*}
with 5 free real parameters $c_i$. To obtain $u^3$ we have several possibilities. One is to take
\begin{gather} \label{2.12a}
\left(B_1+\sum_{i=2}^6 c_i B_i \right)^{3/2}= h^3k^3u^3\big[1+ \mathcal O\big(h^2,  hk,  k^2\big)\big].
\end{gather}
The corresponding discrete Liouville equation is then
\begin{gather}
 \alpha[4B_3-(B_4+B_5)]+\beta[4B_6-3(B_4+B_5)]
 =\left(B_1+\sum_{i=2}^6 c_i B_i \right)\!\left\{\left|B_1+\sum_{i=2}^6 c_i B_i \right|\right\}^{1/2}\!\!\!,\!\!\!\label{2.13a}
\end{gather}
with
\begin{gather} \label{2.14a}
2( \alpha-\beta)=1.
\end{gather}

Another possibility is to replace the basis (\ref{2.10a}) by
\begin{gather*} %\label{2.15a}
A_1=B_1, \qquad A_a=B_1+\frac 1 2 B_a, \qquad a=2, \dots, 6.
\end{gather*}
We can then approximate the right-hand side of the discrete algebraic Liouville equation by
\begin{gather*} %\label{2.16a}
\sum_{a,b=1}^6 \gamma_{a,b}   A_a \sqrt{|A_b|} = h^3k^3 u^3  \sum_{a,b=1}^6 \gamma_{a,b}  \big[ 1+ \mathcal O\big(h^2,   hk,  k^2\big)\big].
\end{gather*}
Then the discrete algebraic Liouville equation reads
\begin{gather} \label{2.17a}
 2 \alpha[4A_3-2A_1-(A_4+A_5)]+2 \beta[4A_6+2A_1-3(A_4+A_5)]= \sum_{a,b=1}^6 \gamma_{a,b}   A_a \sqrt{|A_b|}, \\
\sum_{a,b=1}^6 \gamma_{a,b}=2(\alpha-\beta).\nonumber
\end{gather}
The general \looseness=-1 invariant equations (\ref{2.17a}), (\ref{2.12a}) use all 9 points on the stencil and contain a lot of free parameters.
The parameters  can be chosen to simplify calculations,
though the choice of
 $(\alpha, \beta, c_1, \dots, c_6)$ or $(\alpha, \beta, \gamma_{a,b})$ is restricted by the type of boundary conditions we wish to impose.

The quantity $u_{22}$ f\/igures in $J_8$ only. An explicit scheme is obtained if~$J_8$ f\/igures linearly in the corresponding invariant discrete Liouville equation.

One possibility is to choose $\alpha=-3 \beta$ and $c_2=c_3=c_4=c_5=0$, $c_6=1/2$ in (\ref{2.13a}), (\ref{2.14a}). Then $\beta=-1/8$ and the invariant Liouville equation reduces to
\begin{gather} \label{2.18a}
J_8=J_7+3(J_2-J_1)- \frac{1}{\sqrt{2}}(3J_1-J_7)\sqrt{| 3J_1-J_7|}.
\end{gather}
In terms of the f\/ield $u_{ij}$ (\ref{2.18a}) reads
\begin{gather}
u_{22}=\frac{1}{u_{11}} [ u_{12}u_{21}+3(u_{11}u_{00}-u_{10}u_{01})\nonumber\\
\hphantom{u_{22}=}{}
-\frac{1}{\sqrt{2}}hk(3u_{01}u_{10}-u_{12}u_{21})\sqrt{|3u_{01}u_{10}-u_{12}u_{21}|}].\label{2.18b}
\end{gather}
so that $u_{22}$ is expressed in terms of $u_{00}$, $u_{01}$, $u_{10}$, $u_{11}$, $u_{12}$ and $u_{21}$, i.e., only 7 points are involved.

Another simple possibility is to choose $\alpha=-3\beta$ and $\gamma_{ab}=\delta_{a1}\delta_{b6}$ in~(\ref{2.17a}). Then we have  $\beta=-1/8$  and we obtain
\begin{gather} \label{2.19a}
J_8=\frac{J_7+3(J_2-J_1)-\frac 3 {\sqrt{2}} J_1\sign{3J_1-J_7} \sqrt{| 3J_1-J_7|}}{1-\frac 1 {\sqrt{2}} \sign{3J_1-J_7}\sqrt{| 3J_1-J_7|}}
\end{gather}
In terms of the f\/ield $u_{ij}$ (\ref{2.19a}) reads
\begin{gather}
u_{22}=\bigg\{u_{12}u_{21}+3(u_{11}u_{00}-u_{10}u_{01})\nonumber\\
\hphantom{u_{22}=}{} -\frac{3}{\sqrt{2}}hku_{01}u_{10}\sign{3u_{01}u_{10}-u_{12}u_{21}}
\sqrt{|3u_{01}u_{10}-u_{12}u_{21}|}\bigg\}\nonumber\\
\hphantom{u_{22}=}{}
\times \bigg\{u_{11}\bigg[1-\frac{hk}{\sqrt{2}}\sign{3u_{01}u_{10}-u_{12}u_{21}}\sqrt{|3u_{01}u_{10}-u_{12}u_{21}|}\bigg]\bigg\}^{-1}.
\label{2.19b}
\end{gather}
Again only 7 of the 9 points on a stencil are used.

Equations (\ref{2.18b}) and (\ref{2.19b}) are to be viewed as recursion relations, expressing $u_{2,2}$ in terms  of 6 points on a rectangle of which the point (2,2) is the top right
vertex (see Fig.~\ref{f2}).

By construction (\ref{2.18b}) and (\ref{2.19b}) are better approximations of the equation~(\ref{1.1}) than is~(\ref{a2.8}). This does not mean that they will provide better numerical results and some comments are in order.

 1.~Boundary conditions for a numerical solution on a 4-point lattice  require the knowledge of $u(x,y)$ on two lines, e.g., $u_{m,0}$ and $u_{0,n}$, i.e., $u(x,0)$ and $u(0,y)$. On the 9-point lattice we must start with 2 sets of parallel lines, e.g., $u_{m,0}$, $u_{m,1}$ and $u_{0,n}$, $u_{1,n}$. This amounts to giving $u(x,0)$, $u(0,y)$ and the f\/irst term of $u_y(x,0)$, $u_x(0,y)$. This is more information than is needed in standard (non invariant) discretizations and indeed more information than is needed in theory to determine a solution completely. Hence once $u(x,0)$ and $u(0,y)$ are given $u_y(x,0)$ and $u_x(0,y)$ {\it cannot} be chosen arbitrarily. In our numerical solutions we calculated the conditions using an exact solution so $u(x,0)$, $u(0,y)$ and $u_x(0,y)$, $u_y(x,0)$ are consistent.

 2.~Contrary to the case of a 4-point lattice, instabilities close to zero lines of solutions cannot be avoided on 7- or 9-point lattices. Indeed let us give initial conditions on the f\/irst square satisfying $u_{00}\ne 0$,  $u_{01}\ne 0$, $u_{10}\ne 0$, $u_{11}\ne 0$, $u_{12}\ne 0$, $u_{20}=\epsilon_1$, $u_{21}=\epsilon_2$. From the known solution of the PDE~(\ref{1.1}) we expect the solution to satisfy $u_{2,m}=0$ for $m \ge 2$. Equation~(\ref{2.18b}) implies
\begin{gather*} %\label{3.9}
u_{22}=\frac{1}{u_{11}} \bigg[ u_{12} \epsilon_2+3(u_{11}u_{00}-u_{10}u_{01})-\frac{1}{\sqrt{2}}hk(3u_{01}u_{10}-u_{12} \epsilon_2)\sqrt{|3u_{01}u_{10}-u_{12} \epsilon_2|}\bigg].
\end{gather*}
Thus \looseness=-1 $u_{22}$ is not strictly zero for $\epsilon_1=\epsilon_2=0$, it does however satisfy $u_{22} \sim \mathcal O(h^2,k^2,hk)$. This is acceptable, however the problem arises when we shift the stencil and calculate $u_{32}$ which is supposed to be f\/inite and nonzero if we assume $u_{30}\ne 0$, $u_{31}\ne 0$. What we obtain from~(\ref{2.18b})~is
\begin{gather*}
u_{32} = \frac{1}{\epsilon_2} \bigg[ \mathcal O\big(h^2,k^2,hk\big) u_{31}+3(\epsilon_2 u_{10}-\epsilon_1 u_{11}) \\
\hphantom{u_{32} =}{}
-\frac{1}{\sqrt{2}}hk\big(3\epsilon_1 u_{11}-\mathcal O\big(h^2,k^2,hk\big) u_{31}\big)  \sqrt{\big|3\epsilon_1 u_{11}-\mathcal O\big(h^2,k^2,hk\big) u_{31}\big|}\bigg].
\end{gather*}
Thus, $u_{32}$ is singular for $\epsilon_2=0$ and becomes f\/inite only in the continuous limit $h=k=0$. This will quite obviously create numerical instabilities. They are avoided only for very special initial conditions, such that $u_{22}=0$ for all~$h$ and~$k$. Using~(\ref{2.19b}) leads to the same kind of problems.

Sadly (for the 9 points scheme) the answers to all three questions posed in the beginning of this section are negative.

 1.~The only $ {\rm VIR}(x)\otimes {\rm VIR}(y)$ invariant lattice is given by requiring conditions~(\ref{a2.4}) in all four quadrangles of Fig.~\ref{f2}. Indeed we have
\begin{gather*}
\operatorname{pr} X\big(x^3\big) \xi_a=(x_{11}-x_{00})(x_{10}-x_{01}) \xi_a, \\
 \operatorname{pr} Y\big(y^3\big) \eta_a=(y_{11}-y_{00})(y_{10}-y_{01}) \eta_a,\qquad a=1,2,3,4
\end{gather*}
so (weak) invariance requires $\xi_a=\eta_a=0$ and adding further points does not help. Moreover, no function of $J_1, \dots, J_8$ has the correct continuous limit  and is invariant under ${\rm VIR}(x) \otimes {\rm VIR}(y)$.

2.~We have determined that the invariant~(\ref{xyz}) cannot be discretized in an  $ {\rm SL}_x(2 , \mathbb{R})\otimes {\rm SL}_y(2 , \mathbb{R})$ invariant manner using $J_1, \dots, J_8$ but we do not present the details here since this question is not related to
the Liouville equation.

\begin{figure}[t] \centering
\includegraphics[width=0.5\textwidth]{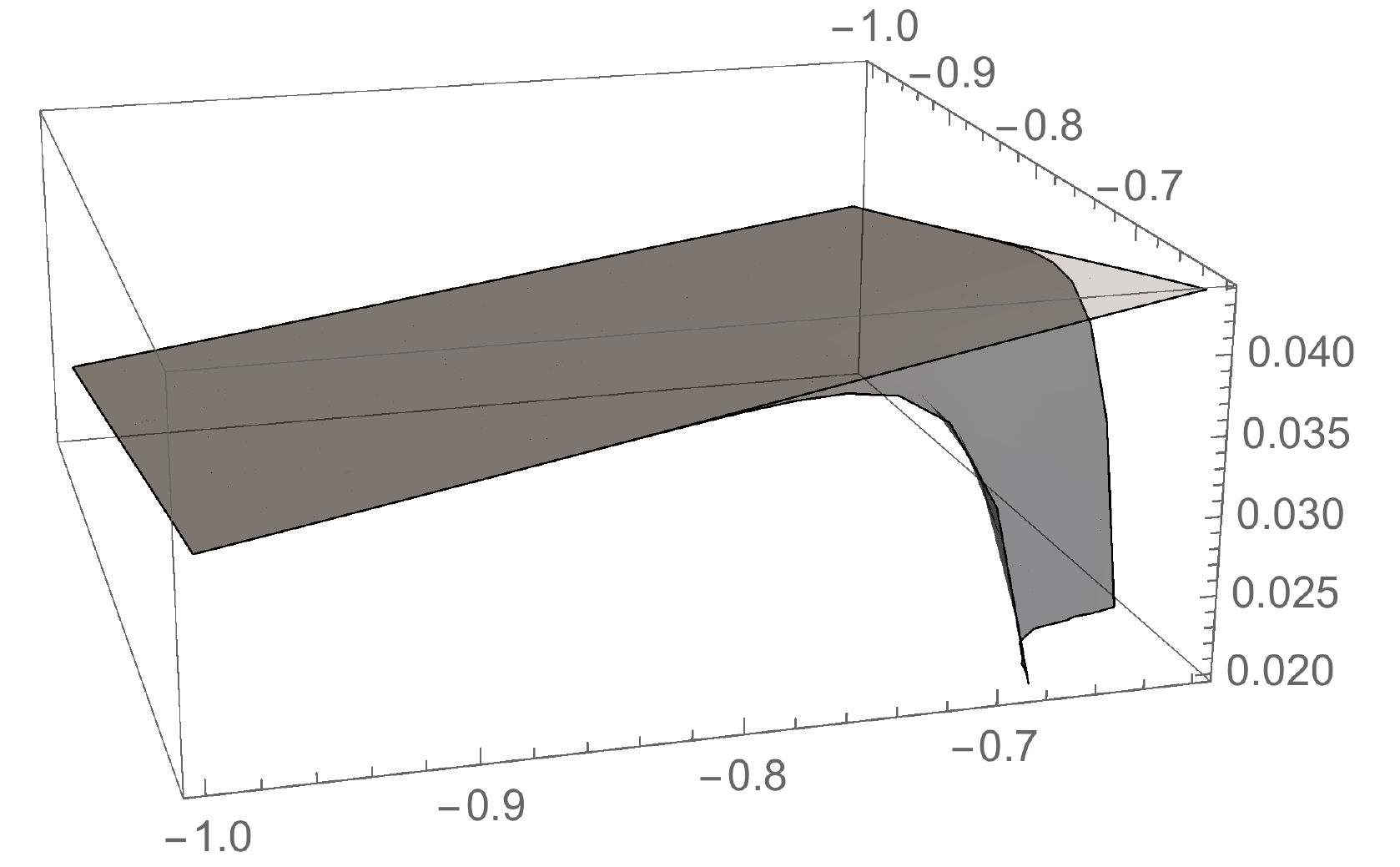}

\caption{Comparison between the 4-point and 7-point approximation for the
solution $f_1$ given in (\ref{f1}). The dark grey graph represents the results of the application of~(\ref{2.19a}),~(\ref{2.19b}), while the light grey one is obtained by using the 4-point formula~(\ref{a2.9}),~(\ref{a2.10}). The integration is carried out starting from the  the point $(-1.0, -1.0)$, the recursion formula is applied for steps $h = k =0.02$ on a grid $20 \times 20$.} \label{f3a}
\end{figure}

 3.~Numerical results for several exact solutions showed that serious instabilities occur for the 7-point scheme. As one can see in Fig.~\ref{f3a} representing the solution~$f_1$ given in~(\ref{f1}), instabilities in
the 7-point case occur almost immediately. The same is true for other
solutions of equation~\eqref{1.2}.

 Our conclusion is that the 9-point (or 7-point) scheme is too
unstable to be useful. We present it here because we think that this
negative result is not a priori obvious and that this discussion may be
useful.

\section[The Adler-Startsev linearizable discrete Liouville equation]{The Adler--Startsev linearizable discrete Liouville equation}\label{section4}

Adler and Startsev \cite{1} have presented a discretization of the algebraic Liouville equation (\ref{1.2}) on a four-point lattice, namely
\begin{gather} \label{Li3}
    a_{m+1,n+1}\left(1+\frac{1}{a_{m+1,n}}\right)\left(1+\frac{1}{a_{m,n+1}}\right)a_{m,n} = 1.
\end{gather}
This equation is linearizable by the substitution
\begin{gather*} %\label{Li3a}
    a_{m,n}= -\frac{(b_{m+1,n}-b_{m,n})(b_{m,n+1}-b_{m,n})}{ b_{m+1,n} b_{m,n+1}},
\end{gather*}
where $b_{m,n}$ satisf\/ies the linear equation
\begin{gather*} %\label{Li3b}
b_{m+1,n+1}-b_{m+1,n}-b_{m,n+1}+b_{m,n}=0.
\end{gather*}
Hence the exact general solution of (\ref{Li3}) is
\begin{gather*} %\label{Li3c}
    a_{m,n}= -\frac{(c_{m+1}-c_m)(k_{n+1}-k_n)}{(c_{m+1}+k_n)(c_m+k_{n+1})},
\end{gather*}
where $c_m$, $k_n$ are arbitrary functions of one index each.

In~\cite{3} we showed, following~\cite{1} that the continuous limit of~(\ref{Li3}), for $a_{m,n}=-\frac{hk}{2} u_{m,n}$ when~$h$ and~$k$ go to zero, gives~(\ref{1.2}) and that it has no continuous point symmetries but must have generalized symmetries.
Moreover by def\/ining $c_m=\phi_1(x_{m,n})$, $k_n=\phi_2(y_{m,n})$ with~$x_{m,n}$ and~$y_{m,n}$ def\/ined in~(\ref{a2.7}) we have
\begin{gather*} %\label{b1.1}
c_{m+1}=\phi_1(x)+h\frac{d \phi_1}{dx}+\mathcal O\big(h^2\big),\qquad
k_{n+1}=\phi_2(y)+k\frac{d \phi_2}{dy}+\mathcal O\big(k^2\big),
\end{gather*}
and thus $a_{m,n}=-hk \frac{\phi_{1,x} \phi_{2,y}}{(\phi_1+\phi_2)^2}+\mathcal O(h^3, h^2 k, h k^2, k^3)$ a f\/irst-order approximation of the general solution of~(\ref{1.2}) given by~(\ref{1.3}) .

\section{Numerical tests of the 4-point scheme}\label{section5}

In this section we shall apply the invariant recursion formula (\ref{a2.9}), (\ref{a2.10})  to solve a set of boundary value problems on a quadrant in the $xy$-plane. Boundary conditions will be given on two orthogonal lines parallel to the~$x$ and~$y$ axes,
respectively, and numerical solutions will be constructed above and to the right of these lines. The numerical solutions will be compared with exact solutions  of the continuous equation for the same boundary conditions. In practice we will start from exact solutions given by choosing $\phi(x,y)=\phi_1(x)+\phi_2(y)$ in~(\ref{1.3}) and calculate the values of these functions on the boundaries. The global estimator which we use is the discrete analog of  relative distance in $L_{\cal D}^2$. We compute the quantity
\begin{gather} \label{5.1}
\chi_{\alpha}( F) = \sqrt{\frac{\sum_{i j}(F_{i j}^{\alpha} - F_{i j})^2}{\sum_{i j} F_{i j}^2}} ,
\end{gather}
where  \looseness=-1 $F_{i j}$  are the values  of the exact solution  $F$ on the lattice sites and  $F_{i j}^{\alpha}$, with $\alpha={\rm Inv}, {\rm AS}, {\rm RV}$, or ${\rm stand}$ are the values computed numerically for the invariant, Adler--Startsev, Rebelo--Vali\-quet\-te or standard discretization, respectively. A~similar analysis is performed for the other recursion formulae. The summation will be over all points of the lattice for which the calculation was performed.

The $\chi_{\alpha}$ quantity~(\ref{5.1}) provides information about the overall averaged behavior of numerical solutions, rather then about their point-by-point behavior. Geometrical features of solutions are better ref\/lected by plots of individual solutions and by  a~relative error function such as
\begin{gather} \label{5.r}
R_{ij}=\left | \frac{F_{i j}^{\alpha}-F_{i j}}{F_{i j}} \right |.
\end{gather}
A characteristic property of all solutions of the algebraic Liouville equation concern the zeroes. They either have no zero in any f\/inite domain~$\mathcal D$, or the zeroes are not isolated, but occur along continuous lines parallel to the~$x$ or~$y$ axes.

We will compare results using four dif\/ferent discretization methods and thus four dif\/ferent recursion formulae, expressing $u_{m+1,n+1}$ in terms of $u_{m,n}$, $u_{m+1,n}$ and $u_{m,n+1}$. For comparison we present the four formulae for the f\/irst position of the stencil, i.e., $m=n=0$. In all cases the left hand side of~(\ref{1.2}) is approximated by
\begin{gather*} %\label{5.2}
u_{11} u_{00} - u_{10} u_{01} = h^2 k^2 ( u u_{xy} - u_x u_y ),
\end{gather*}
where $h$ and $k$ are the lengths of the steps in the $x$ and $y$ directions, respectively. The right-hand side of~(\ref{1.2}) is approximated dif\/ferently in each case. The corresponding recursion formulae  and their continuous limits up to one order beyond the leading one are:

 1.~The {\it invariant} method (\ref{a2.9}), (\ref{a2.10}) (preserving the $ {\rm SL}_x(2 , \mathbb{R})\otimes {\rm SL}_y(2 , \mathbb{R})$ symmetry group as  point symmetries)
\begin{gather} \label{5.3}
  u_{11} u_{00} - u_{10} u_{01}  = hk  [ a u_{01} u_{10}  + (1-a) u_{00} u_{11} ] \sign{u_{01} u_{10}} \sqrt{|u_{01} u_{10} |} ,
 \\ \label{5.3a}
 u u_{xy} - u_x u_y ={u}^{3}+ \big( 3{u}^{2} u_y +
 u_{yy} u_x  + u u_{xyy} \big) \frac{k}{2} +  \big( 3 {u}^{2} u_x + u_y u_{xx} - u u_{xxy}  \big) \frac{h}{2}.
\end{gather}

2.~The Rebelo and Valiquette method (preserving the entire inf\/inite-dimensional symmetry algebra as generalized symmetries)
 \begin{gather} \label{5.4}
   u_{11} u_{00} - u_{10} u_{01}  = hk u_{00} u_{01} u_{10},\\
   %\label{5.4a}
   u u_{xy} - u_x u_y    ={u}^{3}+ \big(  2 {u}^{2}  u_y  +  u_x  u_{yy} - u u_{xyy} \big) \frac{k}{2}+
   \big(  2{u}^{2}  u_x +  u_{xx}  u_y -u u_{xxy}
 \big) \frac{h}{2}.\nonumber
  \end{gather}

3.~The Adler--Startsev method (preserving linearizability of the Liouville equation)
  \begin{gather} \label{5.5}
  u_{11} u_{00} - u_{10} u_{01}  = hk u_{00} u_{11} \left[\frac{u_{01} + u_{10}}{2} - hk u_{01} u_{10} \right],
   \\ \label{5.5a}
 u u_{xy} -
u_x  u_y    ={u}^{3}+ \big(  2 {u}^{2} u_y  +  u_x u_{yy} - u u_{xyy}  \big) \frac{k}{2}+  \big(  2{u}^{2} u_x+ u_{xx}  u_y -u u_{xxy}
 \big) \frac{h}{2}.
 \end{gather}

4.~The {\it standard} method (not preserving any specif\/ic structure) def\/ined on the 4 points of a~square lattice is
   \begin{gather} \label{5.6}
    u_{11} u_{00} - u_{10} u_{01} = hk u_{00}^3, \\
   % \label{5.6a}
   uu_{xy}-u_x u_y  ={u}^{3}+  \left(u_{xx} u_y-u u_{xxy} \right) \frac{h}{2}+
 \left(u_x  u_{yy} -u u_{xyy} \right) \frac{k}{2}.\nonumber
   \end{gather}

Each of these formulae gives a dif\/ferent explicit expression for $u_{11}$ in terms of the already known values of $u_{00}$, $u_{01}$ and $u_{10}$.

Several comments are in order:

 1.~In~(\ref{5.3a}) there is no dependence on the parameter~$a$. It will only appear at the order~$hk$ (not $h^2$ or $k^2$). That is the reason why the dependence of the numerical results depends weakly on the choice of $a$ (see Fig.~\ref{depf2a} below to conf\/irm this).

2.~Formulas (\ref{5.3a}) (invariant method) and (\ref{5.5a}) (Adler--Startsev method) coincide. They dif\/fer in the higher-order terms. Table~\ref{table1}, as could be expected, conf\/irms that the results obtained by these two methods are similar.

\begin{figure}[t]\centering

    a) \includegraphics[width=2.5in]{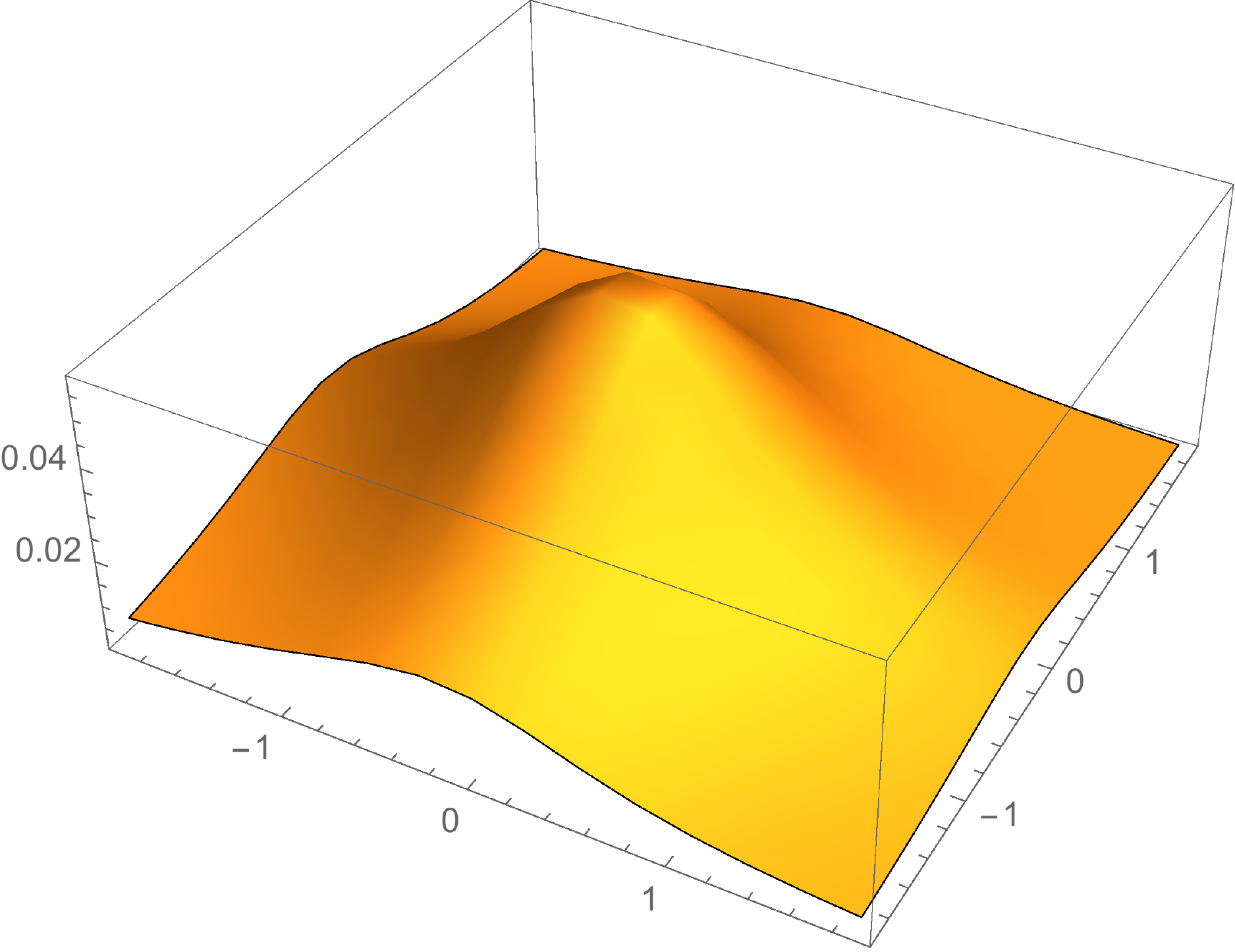}   \qquad b) \includegraphics[width=2.5in]{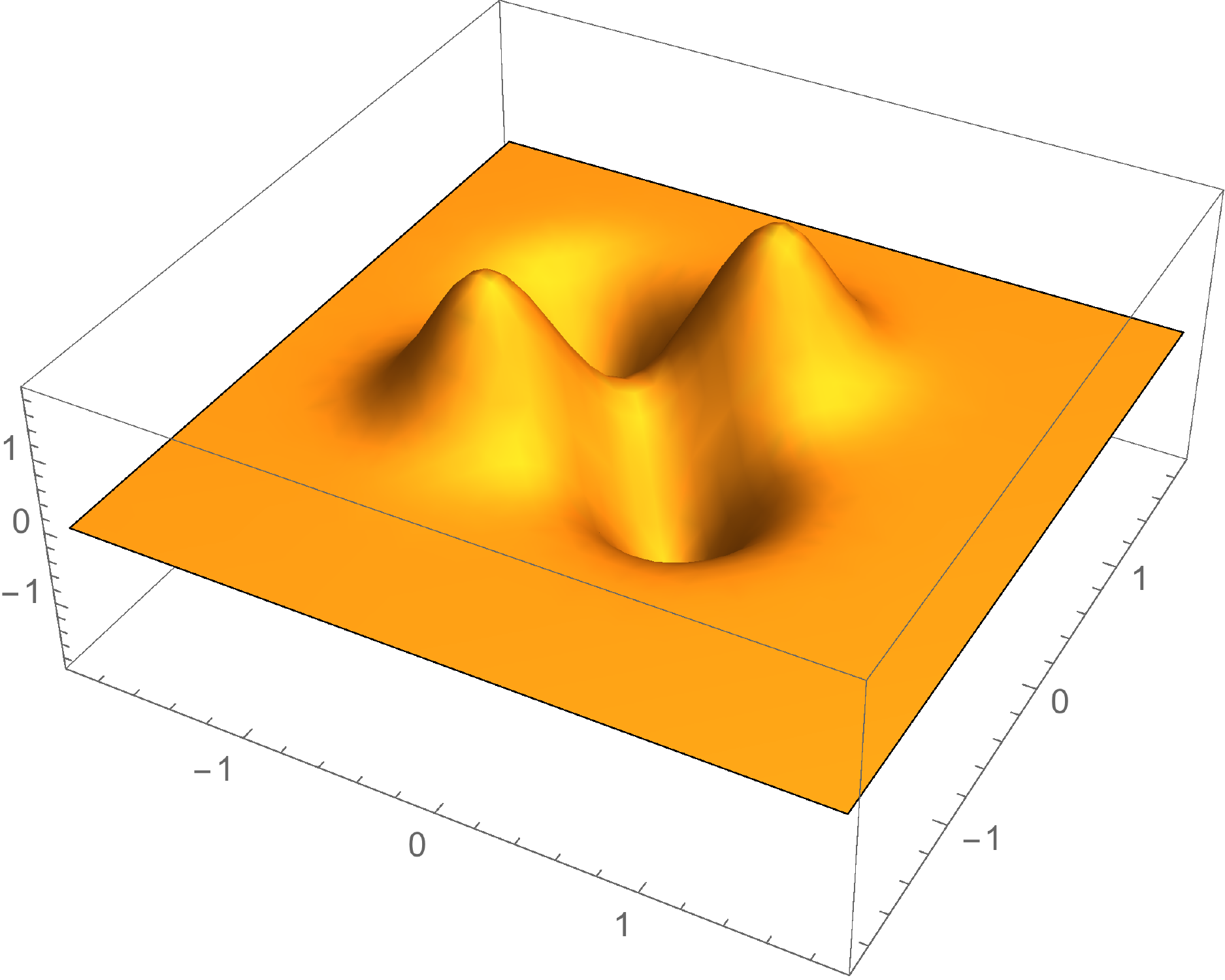}

 c) \includegraphics[width=2.5in]{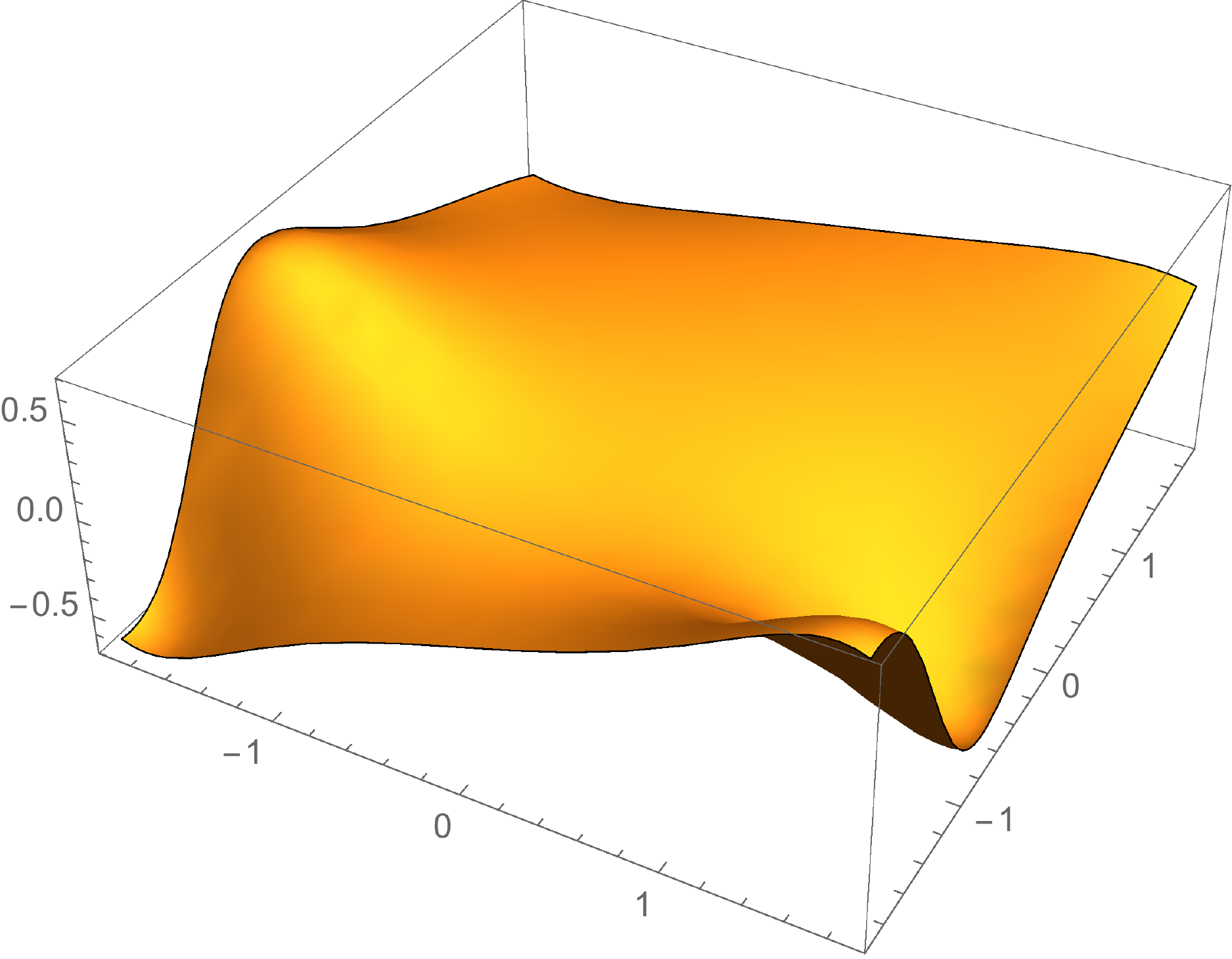} \qquad  d) \includegraphics[width=2.5in]{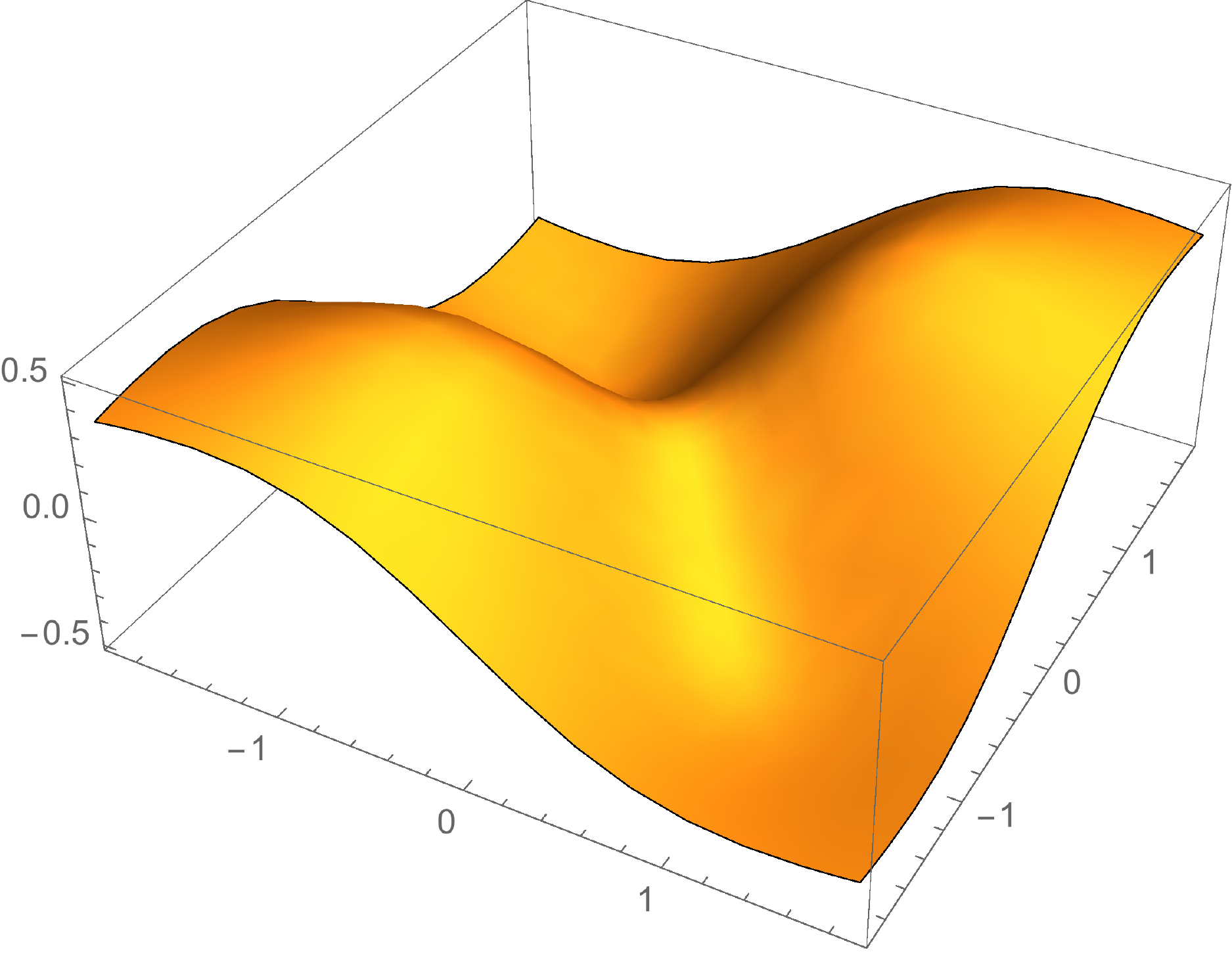}

\caption{Plots of the exact solutions (a) $f_1$, (b) $f_2$, (c) $f_3$  and (d) $f_4$ in the domain ${\cal D}_1 $ used in the numerical integrations.}\label{ff3}
\end{figure}

We consider 5 dif\/ferent solutions of the continuous algebraic Liouville equation (\ref{1.2}), namely
   \begin{gather} % \label{5.7}
   f_1  =   \frac{2   }{\left( x^2+1\right) \left(
   y^2+1\right) \left( \tan ^{-1}(x)+ \tan
   ^{-1}( y)+ 6 \right)^2}, \label{f1}
\\  %\label{5.8}
   f_2 =  \frac{8 \left(1-4 \left(x+\frac{1}{2}\right)\right) (1-4 y)
   \exp \left(-2 x (1+2 x)-2 y(y+2)\right)}{\left(e^{-2 x (1+2 x)}+e^{2 y(1-2 y)}+1\right)^2} ,  \nonumber \\%\label{5.9}
  f_3  =  -\frac{3.38 \sin (1.3 (x+10^{-2})) \cos (1.3 (y+10^{-2}))}{(\cos (1.3 (x+10^{-2}))+\sin (1.3
   (y+10^{-2}))+3)^2},\nonumber \\ %\label{5.11}
   f_4  =  \frac{8 x y}{\left(x^2+y^2+2\right)^2},\nonumber\\
f_5  =  \frac{383.1  e^{3.862 (2.5 (x-0.5)+0.4
   y+2.5)}}{\left(e^{9.655 (x+0.5)}+12.83
   e^{1.545 y}\right)^2}.\nonumber
    \end{gather}
    The functions  $f_1$  and $f_5$ do not contain any zeroes in any f\/inite domain. The functions~$f_2$ and~$f_4$ have one row and one column of zeroes each. The function~$f_3$ contains inf\/initely many  orthogonal lines of zeroes, since it is a periodic function. Finally, $f_5$ is a wall  like function, with no zeroes. We mention that for $f_1$, $f_3$ and $f_4$ the f\/irst-order corrections in~(\ref{5.3a}) and~(\ref{5.5a}) vanish. Plots of the exact solutions $f_1,\dots, f_4$ are given in Fig.~\ref{ff3}, $f_5$ on Fig.~\ref{discrf6}a below.
The right-hand sides of (\ref{5.4}), (\ref{5.5}) and (\ref{5.6}) are polynomials whereas the invariant case~(\ref{5.3}) involves square roots.

 The numerical computations were performed on the square domain ${\cal D}_0 = [-1.5, 1.1] \times [-1.0, 1.6]$, with steps of equal length $h = k = 0.02 $, for a lattice of $130 \times 130$ points. Somewhat arbitrarily we choose the parameter $a$ in the symmetry invariant recursion formula to be $a = 1.0$. The boundary conditions are given on the bottom and left side of the square.

In Table~\ref{table1} we give the $\chi$ quantity (\ref{5.1}) for all f\/ive solutions using 4 dif\/ferent methods.
  \begin{table}[t]\centering
\caption{Relative mean square  distance (\ref{5.1}) between the numerical  solutions and the analytic  one in the domain ${\cal D}_0 $.}\label{table1}
 \begin{tabular}{ l |  c c c r }
 & $\chi_{\rm Inv}$  & $\chi_{\rm AS}$  & $\chi_{\rm RV}$ & $\chi_{\rm stand}$    \\ \hline
  $f_1$ & $5.2 \times 10^{-6}$  & $2.7 \times 10^{-6}$ & $3.1 \times 10^{-4}$ & $ 9.2 \times 10^{-4}$ \tsep{1pt}\\
  $f_2$ & $3.4 \times 10^{-4} $   &   $1.5 \times 10^{-4}$ & $7.6 \times 10^{-3}$& $2.2 \times 10^{-2}$  \\
  $f_3$ & $4.7 \times 10^{-5}$  & $1.5 \times 10^{-5}$ & $3.0 \times 10^{-3}$ & $ 9.2 \times 10^{-3}$ \\
  $f_4$ & $ 4.3 \times 10^{-5}$  & $7.9 \times 10^{-5}$ & $5.2 \times 10^{-3}$ & $ 2.0 \times 10^{-2}$\\
  $f_5$ & $ 3.8 \times 10^{-2}$  & $3.0 \times 10^{-2}$ & $2.8 \times 10^{-1}$ & $ 4.3 \times 10^{-1}$
\end{tabular}
\end{table}
We see from Table~\ref{table1} that $\chi_{\rm Inv}$ and $\chi_{\rm AS}$ are in general of the same order, as are $\chi_{\rm RV}$ and $\chi_{\rm stand}$. The values of  $\chi_{\rm Inv}$ and $\chi_{\rm AS}$ are better than those of $\chi_{\rm RV}$ and $\chi_{\rm stand}$ by at least one order of magnitude, usually by 2~orders, with $\chi_{\rm RV}$ always better than  $\chi_{\rm stand}$. The faster the solution changes, the worse is the result (for all methods), specially for the solutions~$f_2$ and~$f_5$.

  \begin{figure}[t]\centering

  a) \includegraphics[width=2.5in]{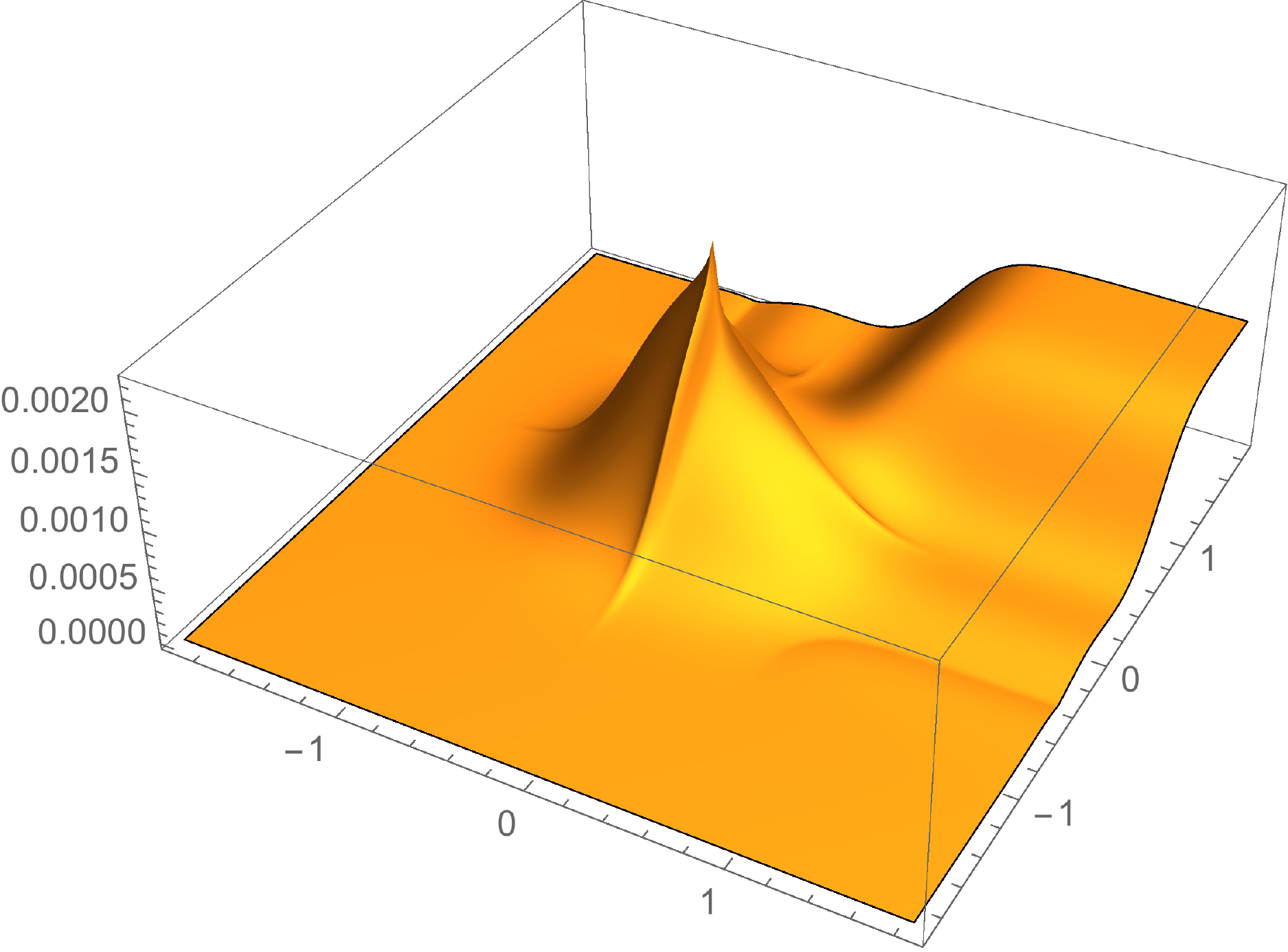}   \qquad  b) \includegraphics[width=2.5in]{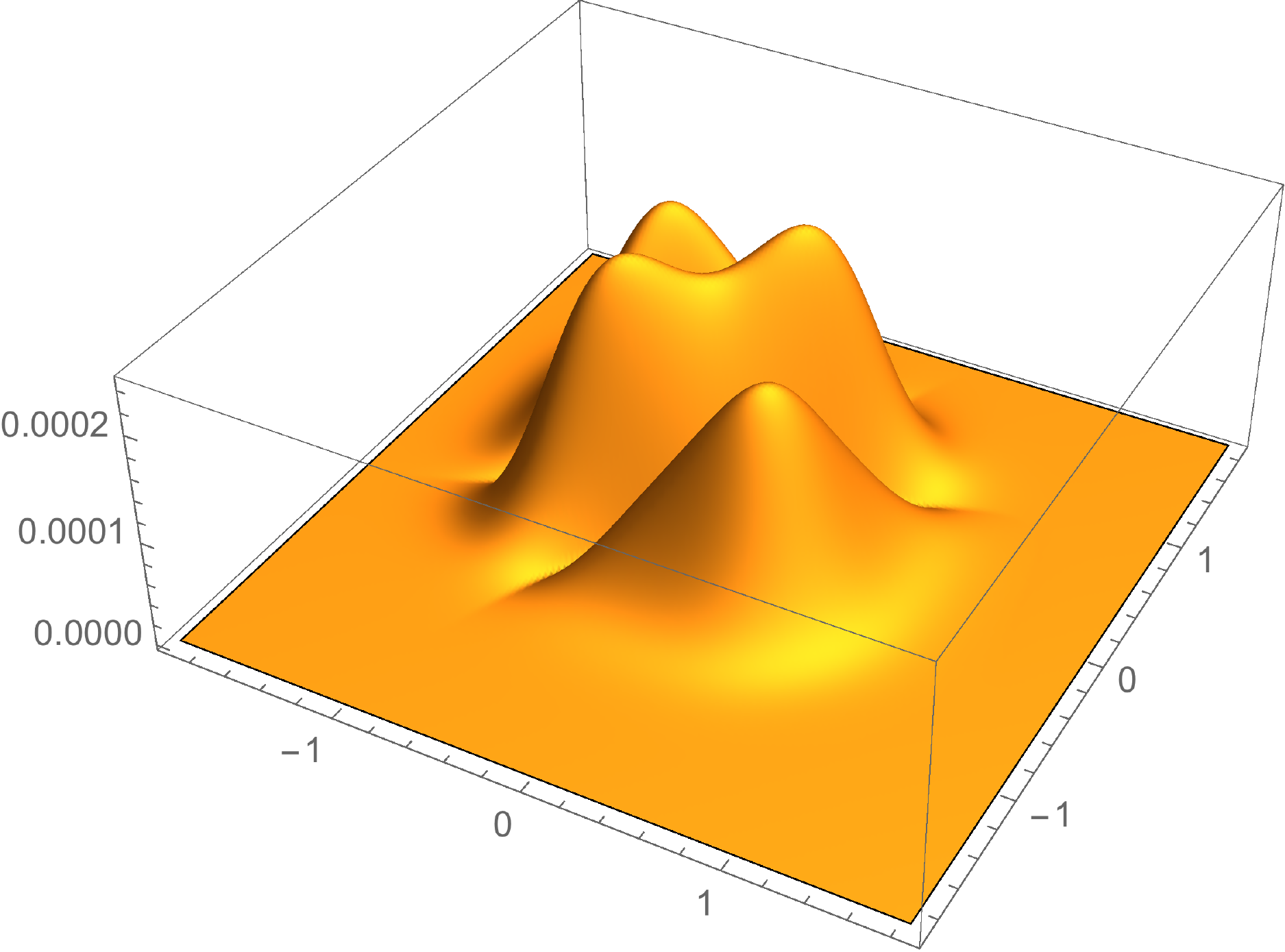}

 c) \includegraphics[width=2.5in]{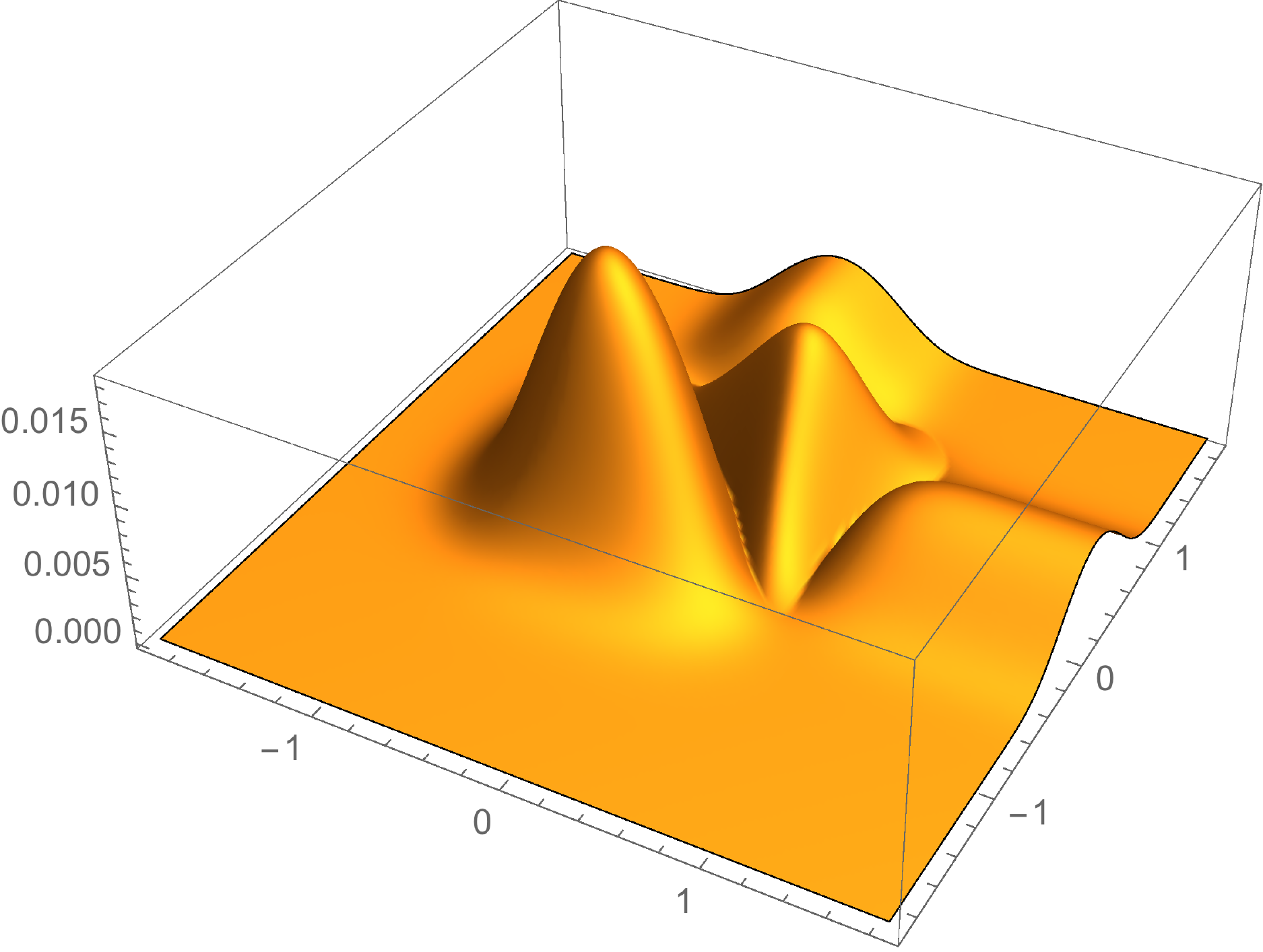} \qquad d) \includegraphics[width=2.5in]{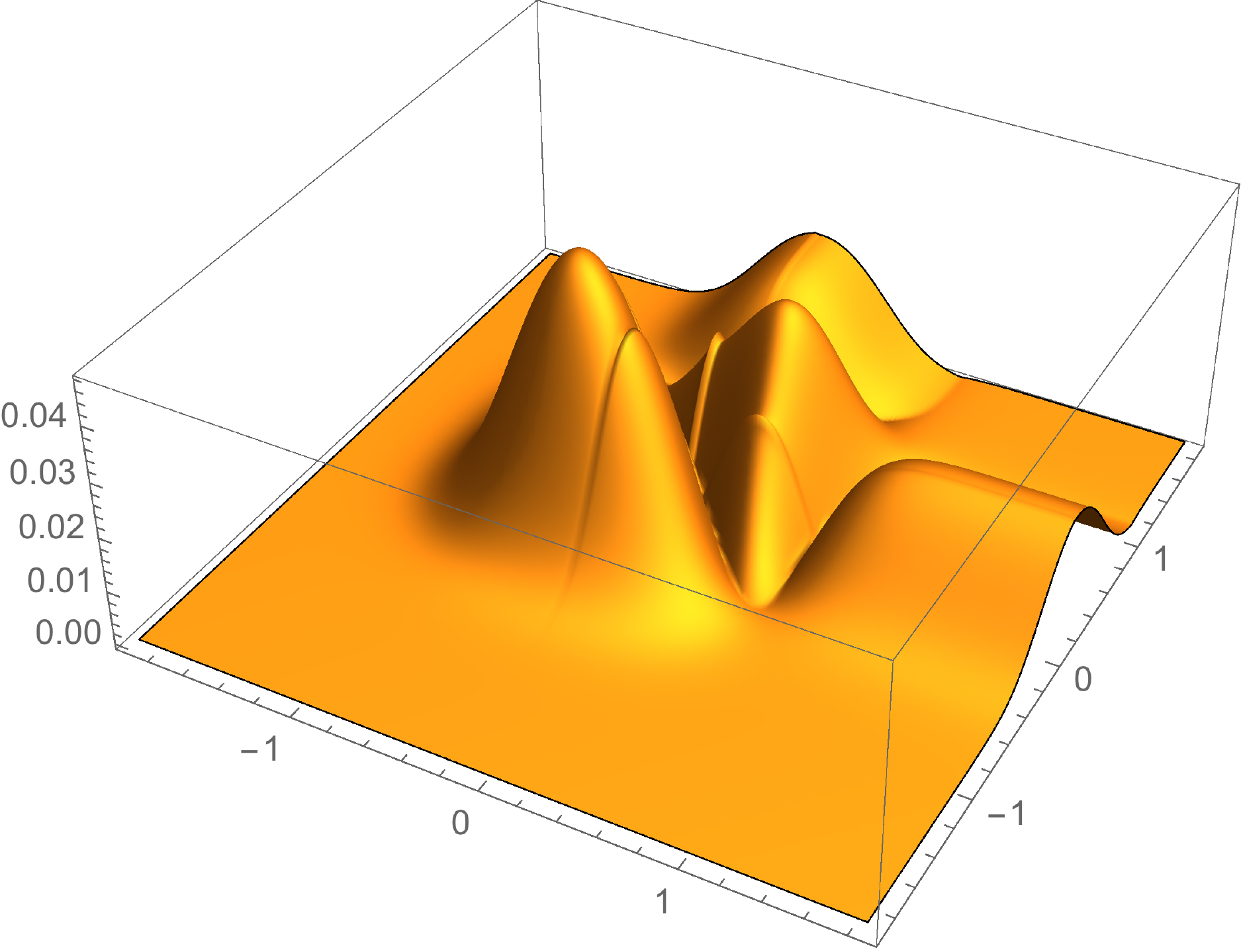}

  \caption{Plots of the relative error $R_{ij}$ def\/ined in~(\ref{5.r}) for the numerical approximation
of the solution~$f_2$ in the domain  ${\cal D}_1$ using the individual
discretizations: (a)~Invariant, (b)~Adler--Startsev,  (c)~Rebelo--Valiquette and (d)~standard.  The graphs are not in the same scale. The maximal
value of the error is approximately $2\times 10^{-3}$, $2\times 10^{-4}$, $1.5\times 10^{-2}$, $4\times 10^{-2}$
for the discretizations (a), (b), (c )  and~(d), respectively.}\label{LocRel}
\end{figure}

 \begin{figure}[t]\centering

 a) \includegraphics[width=2.5in]{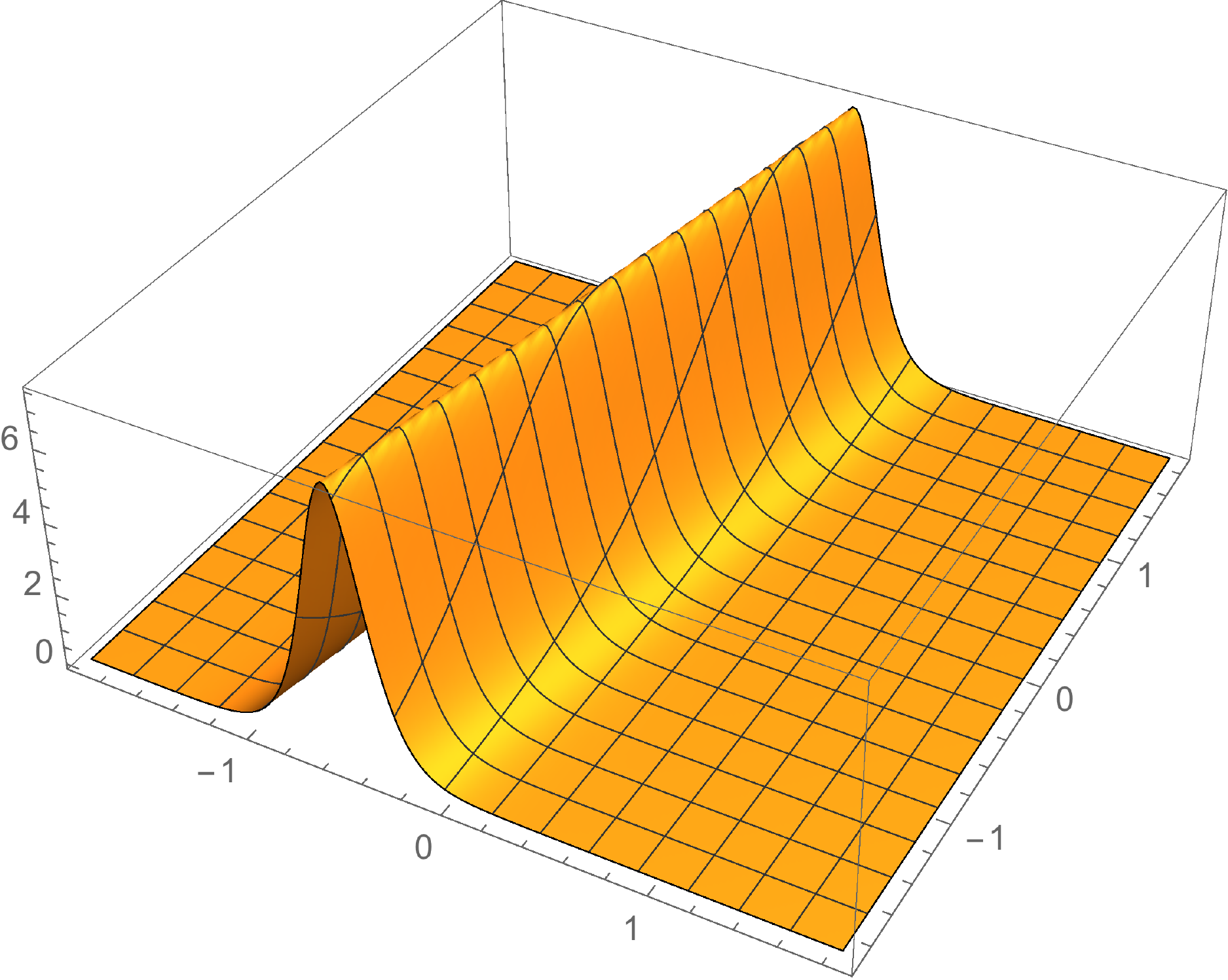}   \qquad  b) \includegraphics[width=2.5in]{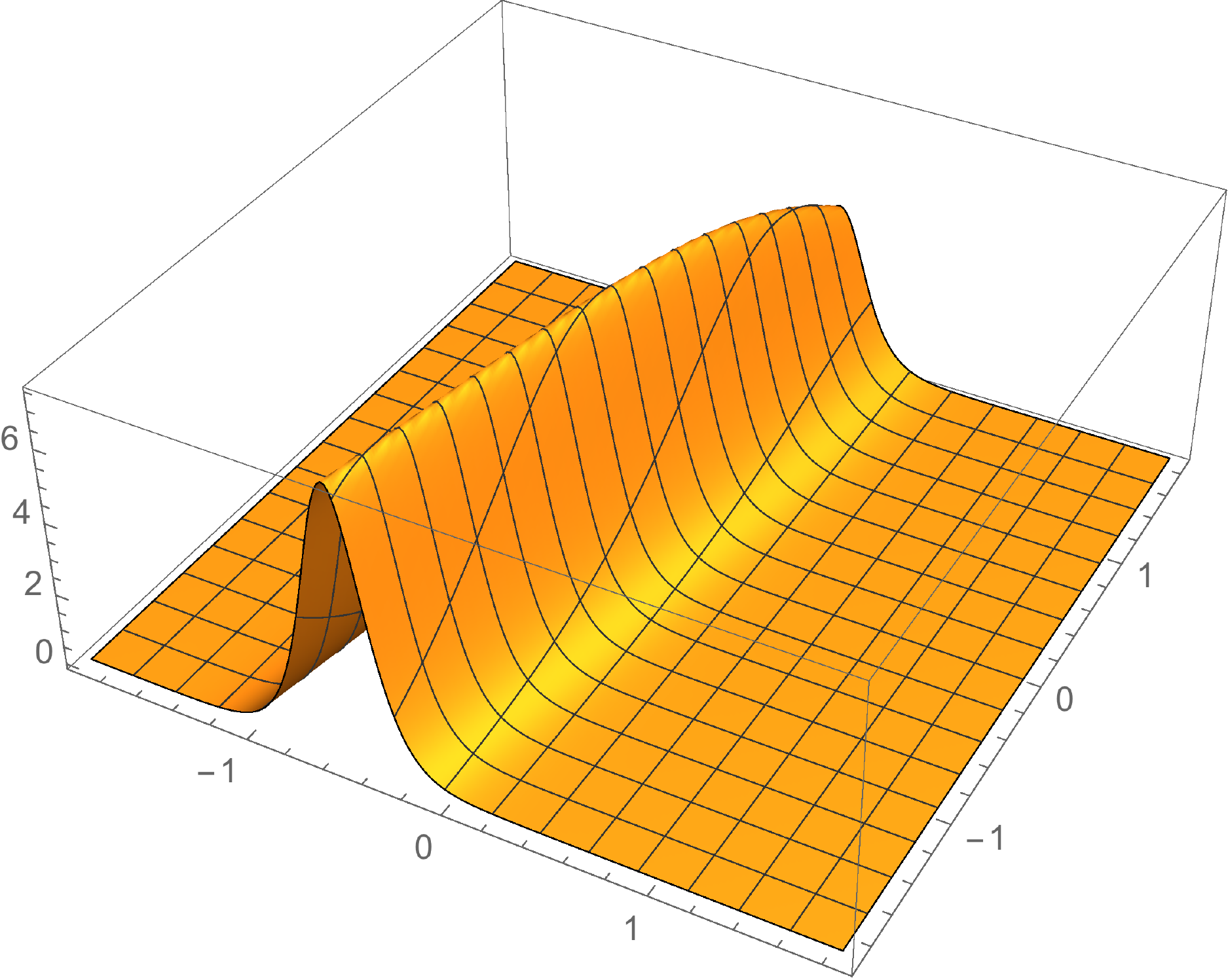}

 c) \includegraphics[width=2.5in]{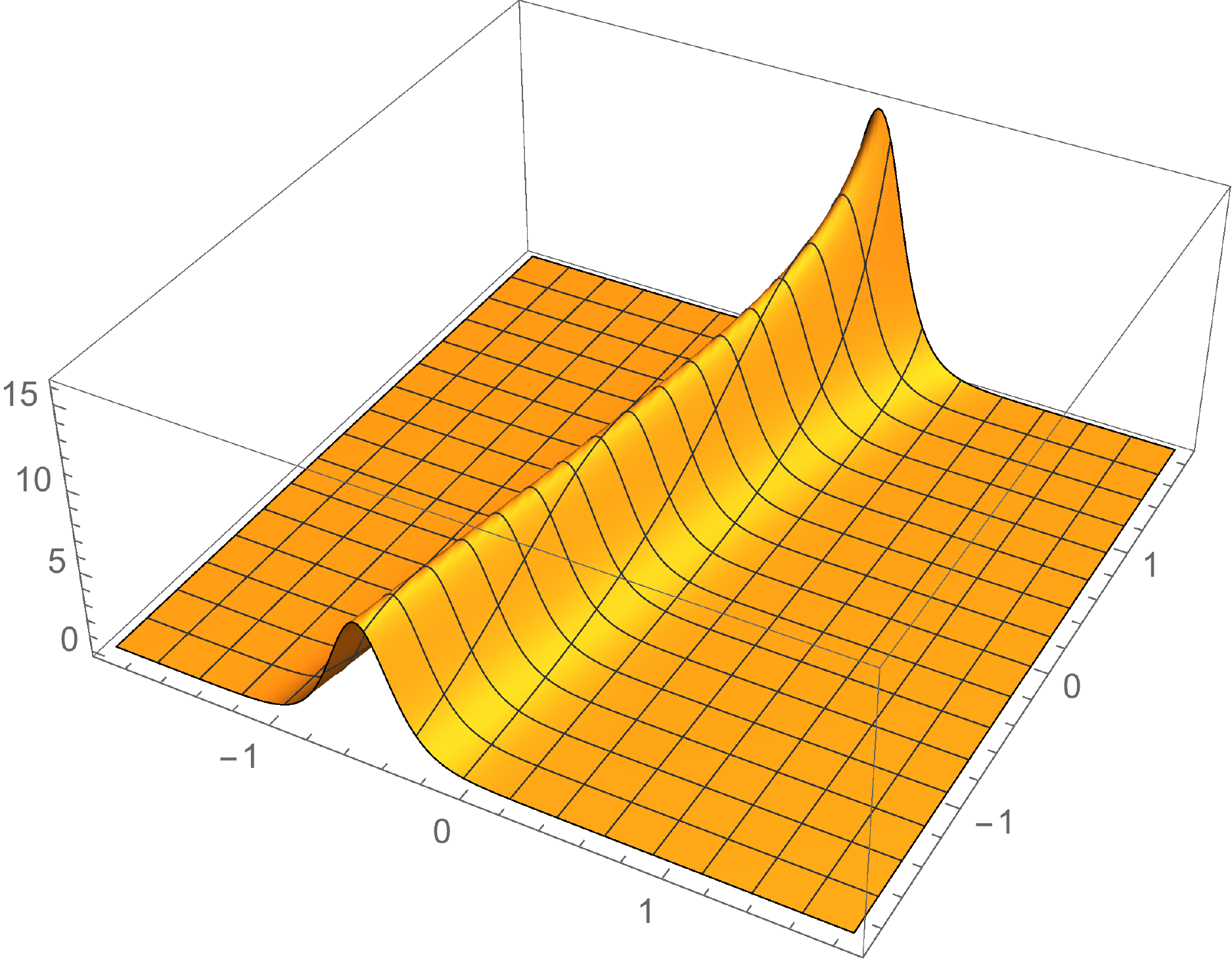} \qquad  d) \includegraphics[width=2.5in]{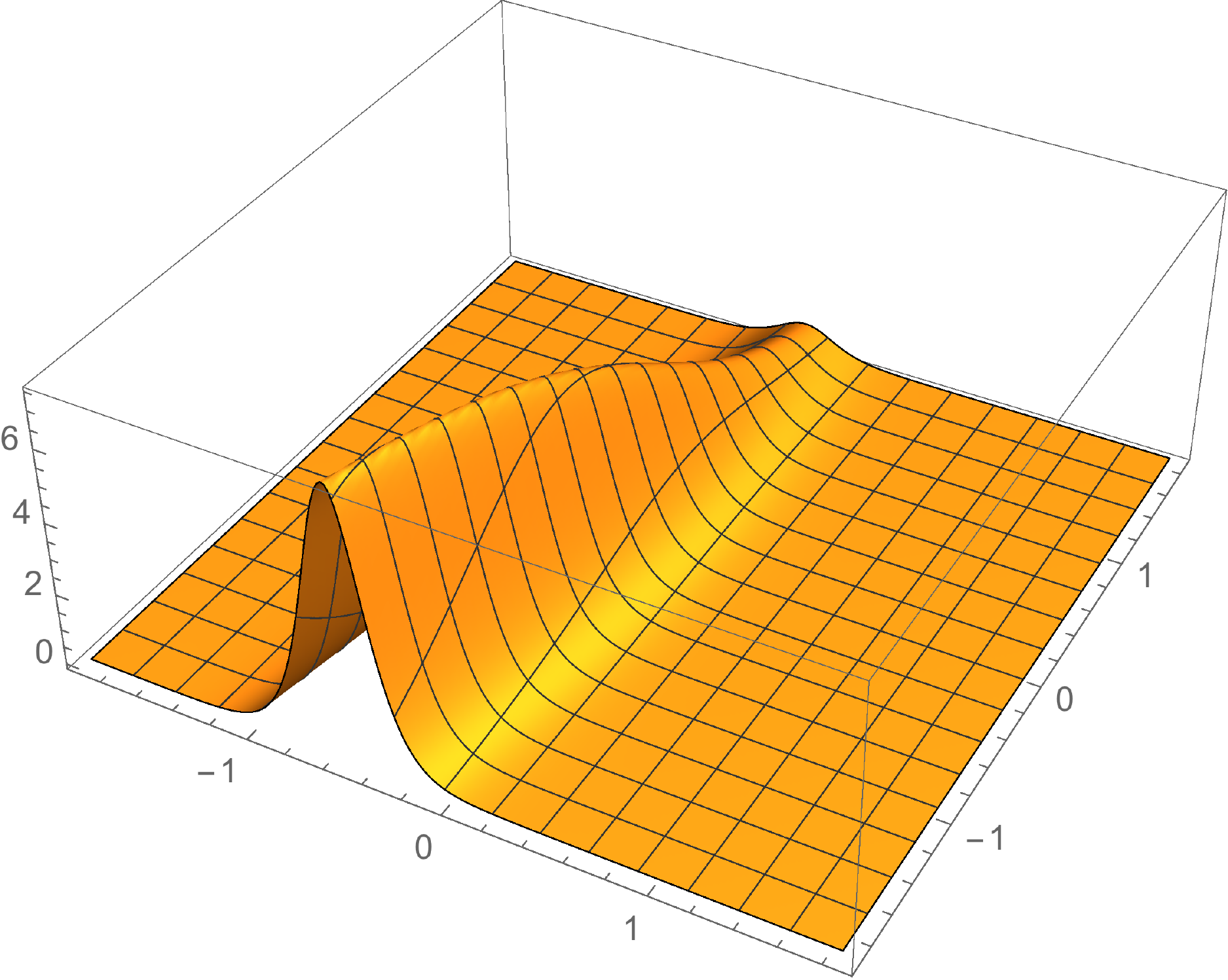}

  e) \includegraphics[width=2.5in]{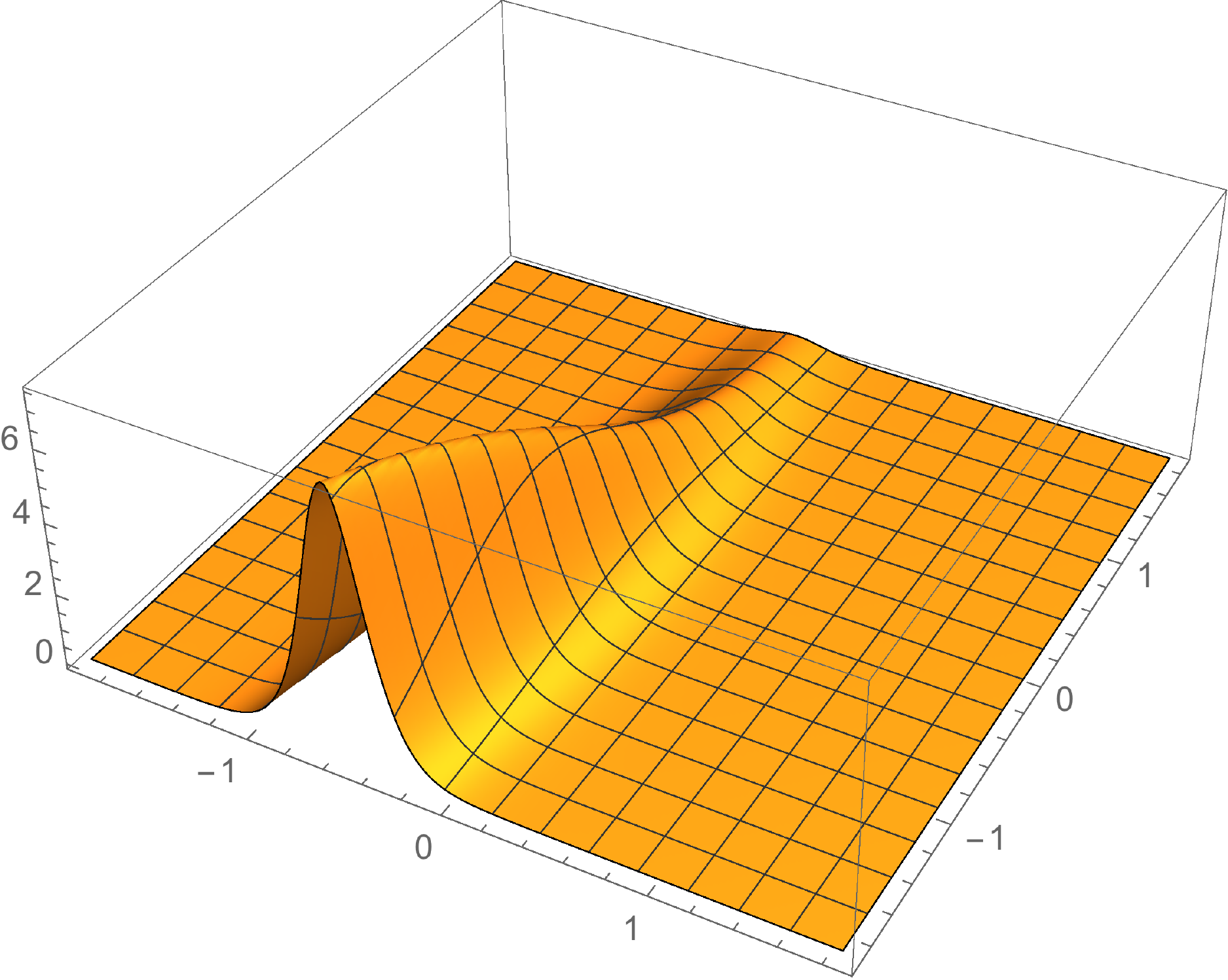}

 \caption{Pictures of the function $f_5$ in the domain ${\cal D}_1$ for the exact solution~(a) and for the individual discretizations: (b)~Invariant, (c)~Adler--Startsev,  (d)~Rebelo--Valiquette and (e)~standard. It is clear by inspection that in this case all proposed discretizations are numerically  unstable, leading to decaying or  blow up of the computed solutions by recursive formulae.}\label{discrf6}
\end{figure}

In order to test the stability of the algorithms with respect to the size of the adopted meshes, we made  another series of calculations involving the above test functions over a f\/ixed  domain  ${\cal D}_1 = [-1.905, 1.895] \times [-1.905, 1.895]$, larger than~${\cal D}_0$, and spanned it using dif\/ferent lattice scales with  $h = k$.
  \begin{table}[t]\centering
\caption{${\rm Log}_{10} \chi$  in the domain ${\cal D}_1$ for each discretization procedure we considered  changing the size of the lattice meshes. We  report only  the results for $f_1$  with the parameter  $a=1$.}\label{table2}
  \begin{tabular}{ c | c  c  c r }
 $h=k$ & $\chi_{\rm Inv}$ &  $\chi_{\rm AS}$  & $\chi_{\rm RV}$ & $\chi_{\rm stand}$     \\ \hline
$8.\times 10^{-1}$ & $-3.98219$  &$-4.30917$  &$-2.57467$   &$-2.13488$       \tsep{1pt}\\
$4.\times 10^{-1  }$   &   $-4.5782$   &$-4.91072$   &$-2.87464$   &$  -2.41647$       \\
$2.\times 10^{-1  }$   &   $-5.17707$   &$-5.51254$   &$-3.17517$   &$-2.7076$       \\
$1.\times 10^{-1  }$   &   $-5.77759$   &$-6.11449$   &$-3.47595$   &$-3.00363$
 \end{tabular}
\end{table}
A general f\/lavor of such calculations can be extracted from Table~\ref{table2}  for the function~$f_1$  For the  solution~$f_1$ (with no zeroes) the value of $\chi_{\rm Inv}$ and $\chi_{\rm AS}$ are comparable  and at least two orders of magnitude lower than the other two. The values of $\chi_{\rm RV}$ are always lower than~$\chi_{\rm stand}$ but of the same order. Generally speaking, decreasing the scale of the mesh by a factor of~$ 0.3$ implies decreasing the value of $\chi$ by a factor of~$ 0.6$ for the invariant and Adler--Startsev discretization and by a factor of~$0.3$ for the other two discretization (quadratic as opposed to linear convergence).

  \begin{table}[t]\centering
\caption{Value of $f_2$ at the four lattice points nearest to the saddle point ($-0.25, +0.25$) for the exact solution and the four numerical approximations.}
\label{table3}
{\footnotesize   \begin{tabular}{ c | c  c  c c r }
point  & Exact &  Inv  & AS   & RV  & stand      \\ \hline
  ($-0.255, 0.255$) &$-4.14391\cdot 10^{-4}$ &$-4.1419\cdot 10^{-4}$  &$-4.14369\cdot 10^{-4}$  & $-4.1281\cdot 10^{-4}$ &$-4.11579\cdot 10^{-4}$  \tsep{1pt}    \\
      ($-0.245, 0.255$) & $4.14391\cdot 10^{-4}$&$4.14257\cdot 10^{-4}$   &$4.14369\cdot 10^{-4}$ &$4.12877\cdot 10^{-4}$& $4.11513\cdot 10^{-4}$     \\
     ($-0.255, 0.245$)&$4.14391\cdot 10^{-4}$ &$4.14123\cdot 10^{-4}$& $4.14369\cdot 10^{-4}$& $4.12744\cdot 10^{-4}$ & $4.11645\cdot 10^{-4}$\\
 ($-0.245, 0.245$)& $-4.14391\cdot 10^{-4}$ & $-4.1419\cdot 10^{-4}$   & $-4.14369\cdot 10^{-4}$ & $-4.1281\cdot 10^{-4}$ & $-4.11579\cdot 10^{-4}$
 \end{tabular}   }
\end{table}

\looseness=1
Let us now turn to the point-by-point behavior of the solutions.  As discussed in Section~\ref{section2} a~cha\-racteristic property of all solutions
of the Liouville equation is that
zeroes are not isolated but occur in straight lines parallel to the axes.
To see how well this is ref\/lected in numerical solutions, let us f\/irst   concentrate on solution $f_2$ which has zeroes on horizontal and vertical lines  passing through the saddle point  $x=-0.25$, $y=+0.25$. These points are not on the lattice due to the def\/inition of the domain~$\mathcal D_1$ with $h=k=0.01$. On Table~\ref{table3} we give the values of the solution $f_2$ at the four lattice points nearest  to the saddle point. The AS solution has the f\/irst 4 digits coinciding with the exact one, the invariant one has~3. The RV and the standard method have~2. Similar results are also valid for the other solutions  with zeroes ($f_3$~and~$f_4$).

The behavior of the solution $f_2$ is plotted on Fig.~\ref{LocRel} for the entire region $\mathcal D_1$  where we show the values of the error function $R_{i,j}$ of~(\ref{5.r}).  The maximal value of $R_{i,j}$ is $2\times 10^{-4}$ in Fig.~\ref{LocRel}b for the AS method, 10 times higher for the invariant method, 75 higher for RV and~200 times higher for the standard one.
For the invariant method the error is concentrated at the saddle point (see Fig.~\ref{LocRel}a) with a~tail behind it.  The AS method has maximal error at the four extrema (see Fig.~\ref{LocRel}b) with no tail.
The other two methods have maximal errors on the maxima of the solutions (not the minima) with tails in both directions.

On Fig.~\ref{discrf6} we analyze the solution~$f_5$ in detail in a point-by-point manner in the domain $\mathcal D_1$.  Fig.~\ref{discrf6}a represents the exact solution (a {\it wall} of constant height). It has no zeroes anywhere in the f\/inite real plane. The height of the wall gradually decreases for solutions~$b$,~$d$ and~$e$, but much more slowly for the invariant method~$b$.  For the AS method the height increases so we present the height of the solution on Fig.~\ref{discrf6}c in a dif\/ferent scale.  The increase in Fig.~\ref{discrf6}c is comparable with the decrease in Fig.~\ref{discrf6}b (a factor of about~2.5).

Finally we analyze the role of the parameter $a$ in the  formula~(\ref{a2.9}),~(\ref{a2.10}). The continuous limit~(\ref{5.3a}) shows that $a$ appears for the f\/irst time in  terms  of order $O(h k)$.  Numerical calculations of~$\chi_{\rm Inv}$ for the function~$f_2$  shows a variation of about a factor~2, when  $a \in [-0.5, 1.5 ]$ (see Fig.~\ref{depf2a}). Furthermore, this function takes a minimum for $a = 0.17$. However, this value is strongly dependent on the test function considered.
\begin{figure}[t]\centering \includegraphics[width=4.8in]{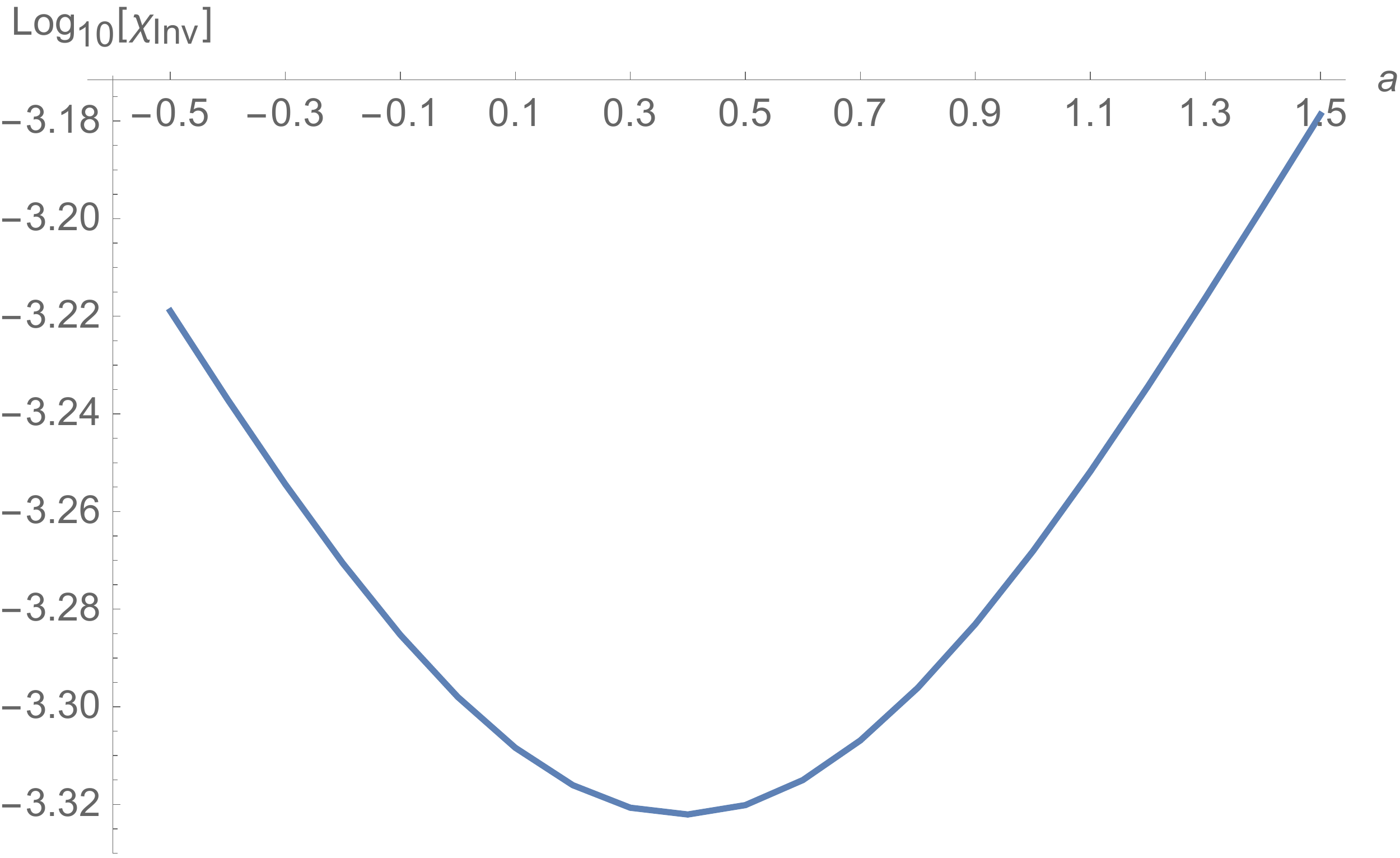}
\caption{Plot of  $\chi_{\rm Inv}$ for the function $f_2$  in the  parameter range $a \in [-0.5, 1.5]$.  } \label{depf2a}
\end{figure}

\section{Conclusions}\label{section6}

Both from the point of physics and from the point of view of geometric integration we  see that for discretizing the Liouville equation we have to choose which characteristic feature of the equation we wish to preserve.  Adler and Startsev  \cite{1} have shown how to preserve linearizability and the existence of  a~class of exact solutions depending on two arbitrary functions  of one variable. We have shown that for a wide class of solutions a recurrence formula based on their method provides the most
accurate results, both using the global $\chi^2$ criterion and comparing
local point by point convergence using the~$R_{ij}$ criterion. On the other hand, linearizability, just like integrability is a property of a very restricted class of nonlinear PDEs.

   The existence of a nontrivial Lie point symmetry group is a much more generic property, specially for PDEs coming from fundamental physical theories. From this point of view the Liouville equation is again special: its Lie point symmetry group is inf\/inite-dimensional.  Rebelo and Valiquette~\cite{2} have presented a discretization that preserves this entire inf\/inite-dimensional symmetry group as a special type of generalized symmetries. As opposed to more general higher symmetries, their symmetries  have a global group action and are very interesting from the theoretical point of view.  From the numerical point of view of we have shown that that the
precision of the RV solutions is systematically better than that of
those obtained by the standard method (though of the same order of
magnitude). The measure of the validity is given by the quantity~$\chi$ of~(\ref{5.1}).

  Finally, the method proposed in~\cite{3} and further developed in this article preserves point invariance under the maximal f\/inite subgroup of the inf\/inite-dimensional symmetry group.  Numerical methods based on this {\it partial} preservation of symmetries perform very well for all solutions in some case even better than the Adler--Startsev case.

  In future work we plan to study symmetry preserving discretizations of other equations with inf\/inite-dimensional symmetry groups, such as the Kadomtsev--Petviashvili equation,  and the three-wave interaction equation.  A symmetry preserving
discretization of the Korteweg--de Vries equation has provided encouraging results~\cite{bihlo}.

\subsection*{Acknowledgments}
DL  has been partly supported by the Italian Ministry of Education and Research, 2010 PRIN {\it Continuous and discrete nonlinear integrable evolutions: from water waves to symplectic maps}.

LM has been partly supported by the Italian Ministry of Education and Research, 2011 PRIN  {\it Teorie geometriche e analitiche dei sistemi Hamiltoniani in dimensioni f\/inite e inf\/inite}. DL and LM are supported also  by INFN   IS-CSN4 {\it Mathematical Methods of Nonlinear Physics}.

 The research of PW is partially supported  by a research grant from NSERC of Canada.
 PW thanks the European Union Research Executive Agency for the award of a Marie Curie International Incoming Research Fellowship  making his stay at University Roma Tre possible. He  thanks the Department of Mathematics and Physics of Roma Tre for hospitality.

We thank the referees for many valuable comments which allowed us to greatly improve the article.

%\cite{LWY2002,15}

\pdfbookmark[1]{References}{ref}
\LastPageEnding

\end{document}